%% file: arxiv.tex
\documentclass[letterpaper,twocolumn,10pt]{article}
\usepackage{usenix2019_v3}

\input{preamble}

\input{arxiv_data_manifest}
\hypersetup{
  pdftitle={\papertitle},
  pdfauthor={Abdulrahman Diaa, Jonathan Petit, Florian Kerschbaum}
}

\begin{document}

\date{}

\title{\Large \bf \papertitle}

\author{
{\rm Abdulrahman Diaa}\\
University of Waterloo
\and
{\rm Jonathan Petit}\\
Qualcomm
\and
{\rm Florian Kerschbaum}\\
University of Waterloo
}

\maketitle
\input{sections/abstract_arxiv}
\input{sections/intro}
\input{sections/background}
\input{sections/problem}
\input{sections/attacks}
\input{sections/method}
\input{sections/evaluation}
\input{sections/discussion}
\input{sections/related_work}
\input{sections/conclusion}

\appendix

\bibliographystyle{plain}
\bibliography{references}
\section{Supplementary Evaluation}
\input{sections/appendix_extra_figures_tables}

\input{sections/appendix_evaluation}

\end{document}

%% file: preamble.tex
\usepackage{xspace}         % smart trailing space after name macros
\usepackage{etoolbox}       % robust string switch for paper-wide result metric

\usepackage{amsmath}        % equations
\usepackage{amssymb}        % math symbols
\usepackage{amsfonts}       % blackboard math symbols
\usepackage{amsthm}         % theorem/definition environments
\usepackage{nicefrac}       % compact symbols for 1/2, etc.

\usepackage{booktabs}       % professional-quality tables
\usepackage{multirow}       % vertically centered headers spanning table rows
\usepackage[inline]{enumitem} % block and inline list spacing controls
\usepackage{graphicx}       % figures
\usepackage{subcaption}     % independently referenceable figure panels
\usepackage{listings}       % wrapped literal prompts in the appendices
\usepackage{pgfplots}       % data-driven figures from csvs/*.csv
\newcommand{\dataplot}{\addplot}
\usepgfplotslibrary{groupplots}
\usepgfplotslibrary{fillbetween}  % interquartile bands in fig:transfer
\pgfplotsset{compat=1.18}
\usetikzlibrary{calc,decorations.pathreplacing,positioning}  % hand-drawn construction figures
\input{plots/data/constants} % generated: bar grid, provider budget, ladder families
\input{tables/data/costs}    % generated: tab:costs rows
\input{plots/data/cross_scheme/constants} % generated: no-box headline results

\input{plots/data/whitebox/constants} % generated: white-box headline results
\input{plots/data/transfer/constants} % generated: transfer sample sizes and ratios
\input{plots/axes}         % shared axis appearance and unit formatting

\input{plots/channels}      % shared attack-result grammar for Eval I
\input{plots/rungs}         % shared ladder grammar for the evaluation ladder
\input{plots/schemes}       % shared per-scheme grammar for the teaser and Eval I
\input{plots/transfer}      % shared grammar for the encoder-transfer panels
\input{plots/channel_response} % shared three-panel channel instrument
\input{plots/attack_channels} % shared grammar for appendix attack trajectories

\definecolor{promptTint}{HTML}{F3F6FA}
\definecolor{promptAccent}{HTML}{315D8A}
\definecolor{exampleGreen}{HTML}{00512E}
\definecolor{exampleYellow}{HTML}{966400}
\definecolor{exampleRed}{HTML}{52000F}
\definecolor{exampleSeed}{HTML}{767676}
\lstdefinestyle{promptblock}{
  basicstyle=\ttfamily\footnotesize\color{black!80},
  backgroundcolor=\color{promptTint},
  frame=single,
  framerule=0.5pt,
  rulecolor=\color{promptAccent!70},
  framesep=5pt,
  aboveskip=0.6\baselineskip,
  belowskip=0.6\baselineskip,
  breaklines=true,
  breakatwhitespace=true,
  breakautoindent=false,
  breakindent=0pt,
  columns=fullflexible,
  keepspaces=true,
  showstringspaces=false,
}

\usepackage{algorithm}      % float wrapper for algorithm listings
\usepackage{algpseudocode}  % algorithmicx pseudocode (\State, \For, \Require)
\makeatletter
\renewcommand{\theHALG@line}{\thealgorithm.\arabic{ALG@line}}
\makeatother

\usepackage{cleveref}       % typed refs: \Cref sentence-start, \cref mid-sentence

\theoremstyle{definition}

\input{commands}

%% file: plots/data/constants.tex
\newcommand{\nprov}{64}
\newcommand{\qbarmin}{65\%}
\newcommand{\qbarmax}{95\%}
\newcommand{\qbarstep}{5\%}
\newcommand{\qbaraxismin}{0.63}
\newcommand{\qbaraxismax}{0.97}
\newcommand{\qbarticks}{0.65,0.75,0.85,0.95}
\newcommand{\rungprefix}{k}
\newcommand{\ladderbase}{\ksemstamp}
\newcommand{\ladderfinal}{\kourstamp}
\newcommand{\altrungprefix}{}
\newcommand{\altladderbase}{\semstamp}
\newcommand{\altladderfinal}{\ourstamp}
\newcommand{\ndocs}{1\,024}
\newcommand{\nnull}{1\,024}

%% file: tables/data/costs.tex
\newcommand{\costasrqualitybar}{90\%}
\newcommand{\resultQualityRequirement}{90\%}
\newcommand{\resultPrimaryFpr}{5\%}
\newcommand{\resultSemEdasAsr}{47.9\%}
\newcommand{\resultKSemEdasAsr}{33.3\%}

\newcommand{\resultSemFidelity}{80.4\%}
\newcommand{\resultKSemFidelity}{80.6\%}
\newcommand{\resultSwordFidelity}{78.6\%}
\newcommand{\resultKSwordFidelity}{76.2\%}
\newcommand{\resultSemCleanTpr}{93.1\%}
\newcommand{\resultKSemCleanTpr}{97.1\%}
\newcommand{\resultSamarkCleanTpr}{86.2\%}
\newcommand{\resultSamarkCleanTprOne}{45.7\%}
\newcommand{\resultSwordCleanTpr}{94.9\%}
\newcommand{\resultKSwordCleanTpr}{97.0\%}
\newcommand{\resultPriorEdapAsrMin}{32.6\%}
\newcommand{\resultPriorEdapAsrMax}{47.9\%}
\newcommand{\resultSamarkDipperAsr}{18.7\%}
\newcommand{\resultSwordEdasAsrReduction}{29.2}
\newcommand{\resultKSwordEdasAsrReduction}{22.5}
\newcommand{\resultSwordFidelityCost}{1.8}
\newcommand{\resultKSwordFidelityCost}{4.4}

\newcommand{\resultSamarkThreshold}{7.02}
\newcommand{\resultPriorEdapAsrRange}{\resultPriorEdapAsrMin--\resultPriorEdapAsrMax}
\newcommand{\publishedcostrows}{%
\samark & \edaconfigbagsent & \underline{45.7} / \underline{86.2} & 13.53 / 7.02 & \underline{44.5} / \underline{15.6} & 45.0 / 18.7 & \underline{61.3} / 32.6 & 81.5 [80.2,82.8] & 64.0 & \underline{6.92} & 2.40 & \textbf{2.5} & \textbf{0.3} \\
\pmark\ (online) & \edaconfigpossent & 95.6 / 97.2 & 2.33 / 1.64 & 5.2 / 2.8 & 40.3 / 26.9 & 57.8 / 46.1 & \textbf{86.0 [84.8,87.2]} & 64.0 & 6.26 & 2.54 & 3.6 & 0.6 \\
}
\newcommand{\lshladdercostrows}{%
\semstamp & \edaconfigpossent & 84.7 / 93.1 & 2.84 / 2.12 & 19.1 / 11.9 & \underline{46.7} / \underline{39.7} & 57.7 / \underline{47.9} & 80.4 [78.9,81.9] & 8.8 & 6.16 & 2.58 & 6.4 & 4.5 \\
\quad\rungaddbestn & \edaconfigpossent & \textbf{97.5} / \textbf{99.3} & 2.84 / 2.12 & 8.5 / 4.9 & 40.4 / 33.4 & 47.9 / 38.1 & 79.6 [78.0,81.2] & 64.0 & 6.06 & 2.48 & 8.6 & 6.7 \\
\quad\rungaddfixed & \edaconfigbagsent & 92.4 / 97.4 & 3.29 / 2.26 & 7.6 / 3.6 & 11.6 / 6.0 & 33.1 / 19.8 & \underline{72.7 [70.9,74.6]} & 64.0 & \textbf{5.35} & 2.48 & \underline{15.7} & \underline{12.4} \\
\quad\rungadddiverse & \edaconfigbagsent & 91.1 / 96.5 & 3.29 / 2.26 & 10.4 / 4.8 & 14.6 / 7.9 & 37.2 / 24.1 & 75.5 [73.8,77.3] & 64.0 & 5.91 & 2.44 & 10.6 & 8.3 \\
\quad\rungfinaltableentry & \edaconfigbagspan & 90.6 / 94.9 & 3.61 / 2.49 & 9.8 / 5.2 & 12.7 / 6.5 & 30.2 / 18.7 & 78.6 [77.1,80.1] & 103.2 & 6.62 & \textbf{3.27} & 6.6 & 3.7 \\
}
\newcommand{\kmeansladdercostrows}{%
\ksemstamp & \edaconfigpossent & 92.9 / 97.1 & 2.84 / 1.76 & 10.4 / 5.2 & 39.4 / 30.3 & 48.4 / 33.3 & 80.6 [79.1,82.1] & \textbf{4.9} & 6.25 & 2.45 & 5.2 & 2.9 \\
\quad\rungaddbestn & \edaconfigpossent & 96.3 / 98.5 & 2.84 / 1.76 & \textbf{4.9} / \textbf{2.1} & 31.2 / 21.9 & 32.6 / 20.0 & 75.2 [73.5,76.8] & 64.0 & 5.86 & 2.42 & 9.4 & 5.9 \\
\quad\rungaddfixed & \edaconfigbagsent & 89.6 / 94.5 & 4.02 / 2.84 & 9.2 / 4.2 & 12.4 / 6.1 & 23.8 / 12.1 & 76.0 [74.5,77.5] & 64.0 & 5.61 & \underline{2.27} & 10.5 & 5.8 \\
\quad\rungadddiverse & \edaconfigbagsent & 86.6 / 92.5 & 4.02 / 2.84 & 14.3 / 7.8 & 21.7 / 10.9 & 39.7 / 22.6 & 78.7 [77.3,80.2] & 64.0 & 6.27 & 2.32 & 6.1 & 3.4 \\
\quad\rungfinaltableentry & \edaconfigbagspan & 93.6 / 97.0 & 3.58 / 2.40 & 6.3 / 3.3 & \textbf{8.7} / \textbf{4.5} & \textbf{20.6} / \textbf{10.8} & 76.2 [74.7,77.8] & \underline{107.4} & 6.75 & 3.25 & 7.0 & 5.0 \\
}

%% file: plots/data/cross_scheme/constants.tex
\newcommand{\resultSwordEdasAsr}{18.7\%}
\newcommand{\resultSwordEdapAsr}{16.3\%}
\newcommand{\resultKSwordEdasAsr}{10.8\%}
\newcommand{\resultKSwordEdapAsr}{8.5\%}
\newcommand{\resultSamarkEdasAsr}{32.6\%}
\newcommand{\resultSamarkEdapAsr}{32.6\%}

%% file: plots/data/whitebox/constants.tex
\newcommand{\resultWhiteboxKSemAsrMin}{52.2\%}
\newcommand{\resultWhiteboxKSemAsrMax}{65.5\%}
\newcommand{\resultWhiteboxKSemAsrRange}{\resultWhiteboxKSemAsrMin--\resultWhiteboxKSemAsrMax}
\newcommand{\resultWhiteboxKSwordAsrMin}{19.4\%}
\newcommand{\resultWhiteboxKSwordAsrMax}{39.7\%}
\newcommand{\resultWhiteboxKSwordAsrRange}{\resultWhiteboxKSwordAsrMin--\resultWhiteboxKSwordAsrMax}
\newcommand{\resultWhiteboxKSemEvasionLowBudget}{70.2\%}
\newcommand{\resultWhiteboxKSwordEvasionLowBudget}{29.6\%}
\newcommand{\resultWhiteboxKSwordEvasionHighBudget}{85.4\%}
\newcommand{\resultWhiteboxKSwordEvadingQualityPassHighBudget}{46.5\%}

%% file: plots/data/transfer/constants.tex
\newcommand{\resultTransferDocs}{1\,024}

%% file: plots/axes.tex
\newlength{\plotpanelheight}
\pgfplotsset{
  paper axis/.style={
    grid=major,
    grid style={line width=0.3pt, draw=black!12},
    axis line style={draw=black!40},
    tick style={draw=black!40},
    ticklabel style={font=\footnotesize, /pgf/number format/fixed},
    label style={font=\small},
  },
  fraction x as percent/.style={
    xticklabel={%
      \pgfmathparse{100*\tick}%
      \pgfmathprintnumber[fixed,precision=0]{\pgfmathresult}%
    },
  },
  fraction y as percent/.style={
    yticklabel={%
      \pgfmathparse{100*\tick}%
      \pgfmathprintnumber[fixed,precision=0]{\pgfmathresult}%
    },
  },
}

%% file: plots/channels.tex
\pgfplotsset{
  discard if not/.style 2 args={
    x filter/.append code={
      \edef\tempa{\thisrow{#1}}
      \edef\tempb{#2}
      \ifx\tempa\tempb\else\def\pgfmathresult{nan}\fi
    }
  },
}
\definecolor{channelSentence}{HTML}{2A78D6}
\definecolor{channelDipper}{HTML}{EDA100}
\definecolor{channelAdaptive}{HTML}{E34948}
\definecolor{channelOracle}{HTML}{6C5AA0}

\pgfplotsset{
  channel frontier group/.style={
    group/group size=3 by 1,
    group/horizontal sep=7pt,
    group/ylabels at=edge left,
    group/yticklabels at=edge left,
  },
  channel signature group/.style={
    group/group size=3 by 1,
    group/horizontal sep=32pt,
  },
  channel base/.style={
    paper axis,
    fraction x as percent,
    width=0.35\textwidth, height=\plotpanelheight,
    title style={font=\small, yshift=-3pt},
    clip mode=individual,
    filter discard warning=false,
    unbounded coords=jump,
  },
  channel tradeoff/.style={
    channel base,
    fraction y as percent,
    xmin=0.68, xmax=0.97,
    ymin=\DetectionMetricYMin,
    enlarge y limits={upper,value=0.10},
    xtick={0.7,0.8,0.9},
  },
  channel structure/.style={
    channel base,
    xmin=0.15, xmax=0.51,
    xtick={0.2,0.3,0.4},
  },
  channel lexical detection/.style={
    channel structure,
    fraction y as percent,
    ymin=\DetectionMetricYMin,
    enlarge y limits={upper,value=0.10},
  },
  channel reorder/.style={
    channel structure,
    fraction y as percent,
    ymin=0, ymax=0.235,
    ytick={0,0.1,0.2},
    ylabel style={xshift=8pt},
  },
  channel reseg/.style={
    channel structure,
    fraction y as percent,
    ymin=0, ymax=0.20,
    ytick={0,0.1,0.2},
    ylabel style={xshift=8pt},
  },
  channel legend/.style={
    legend style={
      at={(0.5,1.23)}, anchor=south,
      legend columns=5, draw=none, font=\footnotesize,
      cells={anchor=west},
      /tikz/every even column/.append style={column sep=5pt},
    },
  },
  channel frontier legend/.style={
    legend style={
      at={(0.5,1.23)}, anchor=south,
      legend columns=3, draw=none, font=\footnotesize,
      cells={anchor=west},
      /tikz/every even column/.append style={column sep=5pt},
    },
  },
  channel/sentence/.style={
    only marks, mark=*, mark size=2.3pt, channelSentence,
  },
  channel/sentence-bigram/.style={
    only marks, mark=o, mark size=2.3pt, channelSentence,
    mark options={fill=white}, line width=0.8pt,
  },
  channel/dipper-order/.style={
    mark=square*, mark size=1.9pt, channelDipper!85!black, line width=0.9pt,
  },
  channel/dipper-lexical/.style={
    mark=diamond, mark size=2.6pt, channelDipper!85!black,
    densely dashed, line width=0.9pt,
  },
  channel/adaptive-surrogate/.style={
    mark=*, mark size=1.9pt, channelAdaptive, line width=0.9pt,
  },
  channel/adaptive-whitebox/.style={
    mark=triangle*, mark size=2.3pt, channelAdaptive!75!black,
    densely dashed, line width=0.9pt,
  },
  frontier/sentence/.style={
    mark=*, mark size=1.7pt, channelSentence, line width=0.9pt,
  },
  frontier/dipper/.style={
    mark=square*, mark size=1.7pt, channelDipper!85!black, line width=0.9pt,
  },
  frontier/adaptive/.style={
    mark=triangle*, mark size=1.9pt, channelAdaptive, line width=1.0pt,
  },
  frontier/adaptive-base/.style={
    frontier/adaptive, mark=triangle, mark size=1.9pt, densely dashed,
    line width=0.8pt, mark options={solid, fill=white, draw=channelAdaptive},
  },
  frontier/oracle-k8/.style={
    mark=diamond*, mark size=1.8pt, channelOracle, densely dashed,
    line width=0.9pt,
  },
  frontier/oracle-k32/.style={
    frontier/oracle-k8, mark=diamond, mark size=1.6pt, line width=0.7pt,
    opacity=0.70, dash pattern=on 1.2pt off 1.2pt,
  },
  frontier/partial/.style={
    densely dashed, opacity=0.48, mark options={opacity=0.48},
  },
  frontier/label/.style={
    channel/inline-label,
    every node near coord/.append style={font=\tiny, fill opacity=0.88},
  },
  frontier/label/sentence/.style={
    frontier/label,
    every node near coord/.append style={
      anchor=north, yshift=-2pt, text=channelSentence!85!black,
    },
  },
  frontier/label/dipper/.style={
    frontier/label,
    every node near coord/.append style={
      anchor=north, yshift=-2pt, text=channelDipper!85!black,
    },
  },
  frontier/label/adaptive/.style={
    frontier/label,
    every node near coord/.append style={
      anchor=south, yshift=2pt, text=channelAdaptive!70!black,
    },
  },
  asr axis/.style={
    paper axis,
    fraction y as percent,
    ymin=0, ymax=1,
    ytick={0,0.2,0.4,0.6,0.8,1.0},
    ylabel={$\ASR$ (\%)},
    error bars/error bar style={line width=0.4pt, draw=black!28},
    error bars/error mark options={rotate=90, mark size=1.5pt,
                                   draw=black!28, line width=0.4pt},
    clip mode=individual,
  },
  qbar axis/.style={
    fraction x as percent,
    xmin=\qbaraxismin, xmax=\qbaraxismax,
    xtick={\qbarticks},
    xlabel={$\qbar$ (\%)},
    ticklabel style={font=\footnotesize},
  },
  channel/inline-label/.style={
    only marks, mark=none, forget plot,
    every node near coord/.append style={
      font=\scriptsize, fill=white, fill opacity=0.90,
      text opacity=1, inner sep=0.35pt,
    },
  },
}

%% file: plots/rungs.tex
\definecolor{rungBase}{HTML}{C1272D}     % the base scheme
\definecolor{rungBestN}{HTML}{E8873A}    % + best-of-N
\definecolor{rungFixed}{HTML}{4A9D9C}    % + fixed valid set
\definecolor{rungDiverse}{HTML}{3D6FB4}  % + diversity
\definecolor{rungSpan}{HTML}{1F3D73}     % + semspans, the final design
\definecolor{rungSpanAlt}{HTML}{7B5EA7}  % the appendix span-policy check

\pgfplotsset{
  rung/base/.style={rungBase,       mark=*,         mark size=1.9pt, line width=1.0pt},
  rung/bestn/.style={rungBestN,     mark=square*,   mark size=1.7pt, line width=0.8pt},
  rung/fixed/.style={rungFixed,     mark=triangle*, mark size=2.1pt, line width=0.8pt},
  rung/diverse/.style={rungDiverse, mark=diamond*,  mark size=2.1pt, line width=0.8pt},
  rung/span/.style={rungSpan,       mark=pentagon*, mark size=2.1pt, line width=1.0pt},
  rung/spanalt/.style={rungSpanAlt, mark=otimes*,   mark size=1.9pt, line width=0.8pt,
                       densely dashed},
  rung/marks/.style={mark options={solid, draw=black!55, line width=0.3pt}},
}

\newcommand{\rungaddbestn}{$+$ best-of-$N$}
\newcommand{\rungaddfixed}{$+$ fixed set}
\newcommand{\rungadddiverse}{$+$ diversity ranking}
\newcommand{\rungaddspan}{$+$ \semspans}
\newcommand{\rungfinalentry}[1]{\rungaddspan\ (\textnormal{#1})}
\newcommand{\rungfinaltableentry}{\rungaddspan\,\swordmark}

\newcommand{\rungentriesfor}[2]{%
  \addlegendimage{rung/base, rung/marks}\addlegendentry{#1}%
  \addlegendimage{rung/bestn, rung/marks}\addlegendentry{\rungaddbestn}%
  \addlegendimage{rung/fixed, rung/marks}\addlegendentry{\rungaddfixed}%
  \addlegendimage{rung/diverse, rung/marks}\addlegendentry{\rungadddiverse}%
  \addlegendimage{rung/span, rung/marks}\addlegendentry{\rungfinalentry{#2}}%
}

%% file: plots/schemes.tex
\definecolor{schSemStamp}{HTML}{B3282B}   % SemStamp, published
\definecolor{schKSemStamp}{HTML}{E08214}  % k-SemStamp
\definecolor{schPMark}{HTML}{7B5EA7}      % PMark, online
\definecolor{schSAMark}{HTML}{5F7A8A}     % SAMark
\definecolor{schOurs}{HTML}{1F3D73}       % FirmStamp, matching rungFour
\definecolor{schKFirmStamp}{HTML}{007C78} % k-FirmStamp

\pgfplotsset{
  scheme/base/.style={line width=0.9pt, mark size=1.9pt,
                      mark options={solid, draw=black!55, line width=0.3pt}},
  scheme/semstamp/.style={scheme/base, schSemStamp,   mark=*},
  scheme/ksemstamp/.style={scheme/base, schKSemStamp, mark=square*},
  scheme/pmark-online/.style={scheme/base, schPMark,  mark=triangle*},
  scheme/samark/.style={scheme/base, schSAMark,       mark=otimes*},
  scheme/firmstamp/.style={scheme/base, schOurs, mark=pentagon*,
                           line width=1.3pt, mark size=2.2pt},
  scheme/kfirmstamp/.style={scheme/base, schKFirmStamp, mark=diamond*,
                            line width=1.3pt, mark size=2.2pt},
  scheme/partial/.style={densely dashed, line width=0.7pt, opacity=0.45,
                         mark size=1.5pt},
}

%% file: plots/transfer.tex
\pgfplotsset{
  transfer/median/.style={no marks, line width=1pt},
  transfer/identity/.style={no marks, black!45, dashed, line width=0.6pt},
  transfer/axis/.style={
    paper axis,
    scale only axis,
    width=0.74\columnwidth, height=\plotpanelheight,
    xlabel={surrogate displacement},
    ylabel={provider displacement},
    xmin=0, xmax=0.8, ymin=0, ymax=1.06,
    xtick={0,0.2,0.4,0.6,0.8},
    ytick={0,0.2,0.4,0.6,0.8,1.0},
    ticklabel style={/pgf/number format/precision=1},
    legend style={
      at={(0.98,0.03)}, anchor=south east,
      draw=none, font=\scriptsize, legend columns=1,
      cells={anchor=west}, fill=white, fill opacity=0.75, text opacity=1,
    },
    clip mode=individual,
  },
}

\newcommand{\transferband}[3]{%
  \dataplot [draw=none, forget plot, name path=#1lower]
    table [x=x, y=lo] {plots/data/transfer/#1-band.dat};
  \dataplot [draw=none, forget plot, name path=#1upper]
    table [x=x, y=hi] {plots/data/transfer/#1-band.dat};
  \addplot [fill=#2, draw=none, fill opacity=0.16, forget plot]
    fill between [of=#1lower and #1upper];
  \dataplot [transfer/median, #2]
    table [x=x, y=y] {plots/data/transfer/#1-band.dat};
  \addlegendentry{#3}%
}

%% file: plots/channel_response.tex
\newcommand{\chanrespprefix}{}

\newcommand{\chanrespplot}[2]{%
  \dataplot [rung/#2, rung/marks, error bars/.cd, y dir=both, y explicit]
    table [x=x, y=y, y error minus=em, y error plus=ep]
    {plots/data/channel_response/#1-\chanrespprefix#2.dat};%
}

\newcommand{\chanrespladder}[1]{%
  \chanrespplot{#1}{base}\chanrespplot{#1}{bestn}\chanrespplot{#1}{fixed}%
  \chanrespplot{#1}{diverse}\chanrespplot{#1}{span}%
}

\pgfplotsset{
  chanresp base/.style={
    paper axis,
    fraction x as percent,
    scale only axis,
    width=0.265\textwidth, height=\plotpanelheight,
    xlabel style={yshift=2pt},
    ymin=0, ymax=110,
    ytick={0,25,50,75,100},
    error bars/error bar style={line width=0.45pt, draw=black!30},
    error bars/error mark options={rotate=90, mark size=1.4pt,
                                   draw=black!30, line width=0.45pt},
    clip mode=individual,
  },
  chanresp group/.style={
    group/group size=3 by 1,
    group/horizontal sep=28pt,
    group/ylabels at=edge left,
    group/yticklabels at=edge left,
  },
  chanresp legend/.style={
    legend style={
      at={(0.5,1.17)}, anchor=south,
      legend columns=5, draw=none, font=\footnotesize,
      cells={anchor=west},
      /tikz/every even column/.append style={column sep=6pt},
    },
  },
}

\newcommand{\chanrespbody}[3]{%
  \edef\chanrespprefix{#1}%
  \begin{tikzpicture}
    \begin{groupplot}[chanresp base, chanresp group]
      \nextgroupplot[
        xlabel={measured rewording (\%)},
        ylabel={$\zret$ (\%)},
        xmin=0, xmax=0.40,
        xtick={0,0.2,0.4},
      ]
      \chanrespladder{reword}

      \nextgroupplot[
        xlabel={measured reordering (\%)},
        xmin=0, xmax=1.04,
        xtick={0,0.5,1.0},
        chanresp legend,
      ]
      \rungentriesfor{#2}{#3}
      \chanrespladder{reorder}

      \nextgroupplot[
        xlabel={measured resegmentation (\%)},
        xmin=0, xmax=0.35,
        xtick={0,0.15,0.30},
      ]
      \chanrespladder{reseg}

    \end{groupplot}
  \end{tikzpicture}%
}

%% file: plots/attack_channels.tex
\definecolor{channelAdaptiveBag}{HTML}{007C78}
\definecolor{channelAdaptiveSpan}{HTML}{1F3D73}

\pgfplotsset{
  attackchannels/base/.style={
    paper axis,
    fraction x as percent,
    scale only axis,
    width=0.265\textwidth, height=\plotpanelheight,
    ticklabel style={/pgf/number format/precision=0},
    xmin=0.15, xmax=0.54,
    xtick={0.2,0.3,0.4,0.5},
    xlabel={measured rewording (\%)},
    clip mode=individual,
    filter discard warning=false,
    unbounded coords=jump,
  },
  attackchannels/reorder/.style={
    attackchannels/base,
    fraction y as percent,
    ymin=0, ymax=0.25,
    ytick={0,0.1,0.2},
    ylabel={measured reordering (\%)},
  },
  attackchannels/reseg/.style={
    attackchannels/base,
    fraction y as percent,
    ymin=0, ymax=0.28,
    ytick={0,0.1,0.2},
    ylabel={measured resegmentation (\%)},
  },
  attackchannels/quality-pass/.style={
    attackchannels/base,
    fraction y as percent,
    ymin=0, ymax=1.00,
    ytick={0,0.25,0.5,0.75,1.0},
    ylabel={quality-pass rate (\%)},
  },
  attackchannels/signature-size/.style={
    width=0.26\textwidth, height=\plotpanelheight,
  },
  attackchannels/retention-base/.style={
    attackchannels/base,
    width=0.25\textwidth,
    ymin=0, ymax=100,
    ytick={0,25,50,75,100},
  },
  attackchannels/retention-reword/.style={
    attackchannels/retention-base,
  },
  attackchannels/retention-reorder/.style={
    attackchannels/retention-base,
    xmin=0, xmax=0.24,
    xtick={0,0.1,0.2},
    xlabel={measured reordering (\%)},
  },
  attackchannels/retention-reseg/.style={
    attackchannels/retention-base,
    xmin=0, xmax=0.24,
    xtick={0,0.1,0.2},
    xlabel={measured resegmentation (\%)},
  },
  attackchannels/signature-group/.style={
    group/group size=3 by 1,
    group/horizontal sep=40pt,
  },
  attackchannels/retention-group/.style={
    group/group size=3 by 4,
    group/horizontal sep=28pt,
    group/vertical sep=7pt,
    group/ylabels at=edge left,
    group/yticklabels at=edge left,
  },
  attackchannels/pegasus/.style={
    only marks, mark=*, mark size=2.1pt, channelSentence,
  },
  attackchannels/pegasus-bigram/.style={
    only marks, mark=o, mark size=2.1pt, channelSentence,
    mark options={fill=white}, line width=0.8pt,
  },
  attackchannels/parrot/.style={
    only marks, mark=square*, mark size=2.0pt, channelSentence,
  },
  attackchannels/parrot-bigram/.style={
    only marks, mark=square, mark size=2.0pt, channelSentence,
    mark options={fill=white}, line width=0.8pt,
  },
  attackchannels/dipper-order/.style={
    mark=square*, mark size=1.8pt, channelDipper!85!black, line width=0.9pt,
  },
  attackchannels/dipper-lexical/.style={
    mark=diamond, mark size=2.4pt, channelDipper!85!black, densely dashed,
    line width=0.9pt,
  },
  attackchannels/adaptive-pos-sent/.style={
    mark=triangle*, mark size=2.0pt, channelAdaptive, line width=0.9pt,
  },
  attackchannels/adaptive-bag-sent/.style={
    mark=diamond*, mark size=2.0pt, channelAdaptiveBag, densely dashed,
    line width=0.9pt,
  },
  attackchannels/adaptive-bag-semspans/.style={
    mark=pentagon*, mark size=2.0pt, channelAdaptiveSpan, densely dashdotted,
    line width=0.9pt,
  },
  attackchannels/adaptive/.style={
    mark=triangle*, mark size=2.0pt, channelAdaptive, line width=0.9pt,
  },
  attackchannels/adaptive-base/.style={
    attackchannels/adaptive, densely dashed,
    mark options={solid, fill=white, draw=channelAdaptive},
  },
  attackchannels/legend/.style={
    legend to name=attack-channel-legend,
    legend style={
      legend columns=6, draw=none, font=\scriptsize,
      cells={anchor=west},
      /tikz/every even column/.append style={column sep=2pt},
    },
  },
  attackchannels/retention-legend/.style={
    legend to name=attack-retention-legend,
    legend style={
      legend columns=5, draw=none, font=\scriptsize,
      cells={anchor=west},
      /tikz/every even column/.append style={column sep=2pt},
    },
  },
  attackchannels/inline-label/.style={
    only marks, mark=none, forget plot,
    every node near coord/.append style={
      font=\tiny, fill=white, fill opacity=0.90,
      text opacity=1, inner sep=0.35pt,
    },
  },
}

\newcommand{\attackchannelevasionplot}[3]{%
  \dataplot [attackchannels/#2, discard if not={series}{#2}]
    table [x=evading_reword, y=evading_#3] {plots/data/attack_channels/#1.dat};%
}

\newcommand{\attackchannelqualitypassplot}[2]{%
  \dataplot [attackchannels/#2, discard if not={series}{#2}]
    table [x=evading_reword, y=evading_quality_pass] {plots/data/attack_channels/#1.dat};%
}

\newcommand{\attackchanneltrajectories}[2]{%
  \attackchannelevasionplot{#1}{pegasus}{#2}%
  \attackchannelevasionplot{#1}{pegasus-bigram}{#2}%
  \attackchannelevasionplot{#1}{parrot}{#2}%
  \attackchannelevasionplot{#1}{parrot-bigram}{#2}%
  \attackchannelevasionplot{#1}{dipper-order}{#2}%
  \attackchannelevasionplot{#1}{dipper-lexical}{#2}%
  \attackchannelevasionplot{#1}{adaptive-pos-sent}{#2}%
  \attackchannelevasionplot{#1}{adaptive-bag-sent}{#2}%
  \attackchannelevasionplot{#1}{adaptive-bag-semspans}{#2}%
}

\newcommand{\attackchannelqualitypasstrajectories}[1]{%
  \attackchannelqualitypassplot{#1}{pegasus}%
  \attackchannelqualitypassplot{#1}{pegasus-bigram}%
  \attackchannelqualitypassplot{#1}{parrot}%
  \attackchannelqualitypassplot{#1}{parrot-bigram}%
  \attackchannelqualitypassplot{#1}{dipper-order}%
  \attackchannelqualitypassplot{#1}{dipper-lexical}%
  \attackchannelqualitypassplot{#1}{adaptive-pos-sent}%
  \attackchannelqualitypassplot{#1}{adaptive-bag-sent}%
  \attackchannelqualitypassplot{#1}{adaptive-bag-semspans}%
}

\newcommand{\attackchannelpointlabel}[4]{%
  \dataplot [attackchannels/inline-label, nodes near coords,
            point meta=explicit symbolic,
            every node near coord/.append style={#4},
            discard if not={series}{#2}]
    table [x=evading_reword, y=evading_#3, meta=setting] {plots/data/attack_channels/#1.dat};%
}

\newcommand{\attackchannelendpointlabel}[5]{%
  \dataplot [attackchannels/inline-label, nodes near coords,
            point meta=explicit symbolic,
            every node near coord/.append style={#5},
            discard if not={series}{#2}, discard if not={endpoint}{#3}]
    table [x=evading_reword, y=evading_#4, meta=setting] {plots/data/attack_channels/#1.dat};%
}

\newcommand{\attackchanneltrajectorylabels}[2]{%
  \attackchannelpointlabel{#1}{pegasus}{#2}{anchor=south east, xshift=-1pt, yshift=1pt, text=channelSentence!85!black}%
  \attackchannelpointlabel{#1}{pegasus-bigram}{#2}{anchor=north west, xshift=1pt, yshift=-1pt, text=channelSentence!85!black}%
  \attackchannelpointlabel{#1}{parrot}{#2}{anchor=south west, xshift=4pt, yshift=3pt, text=channelSentence!85!black}%
  \attackchannelpointlabel{#1}{parrot-bigram}{#2}{anchor=south west, xshift=1pt, yshift=1pt, text=channelSentence!85!black}%
  \attackchannelendpointlabel{#1}{dipper-order}{start}{#2}{anchor=north east, xshift=-1pt, yshift=-1pt, text=channelDipper!85!black}%
  \attackchannelendpointlabel{#1}{dipper-order}{end}{#2}{anchor=south west, xshift=1pt, yshift=1pt, text=channelDipper!85!black}%
  \attackchannelendpointlabel{#1}{dipper-lexical}{start}{#2}{anchor=south west, xshift=1pt, yshift=-1pt, text=channelDipper!85!black}%
  \attackchannelendpointlabel{#1}{dipper-lexical}{end}{#2}{anchor=north east, xshift=-1pt, yshift=1pt, text=channelDipper!85!black}%
}

\newcommand{\attackchannelqualitypasslabels}[1]{%
  \dataplot [attackchannels/inline-label, nodes near coords,
            point meta=explicit symbolic,
            every node near coord/.append style={anchor=south east, xshift=-1pt, yshift=1pt, text=channelSentence!85!black},
            discard if not={series}{pegasus}]
    table [x=evading_reword, y=evading_quality_pass, meta=setting] {plots/data/attack_channels/#1.dat};%
  \dataplot [attackchannels/inline-label, nodes near coords,
            point meta=explicit symbolic,
            every node near coord/.append style={anchor=north west, xshift=1pt, yshift=-1pt, text=channelSentence!85!black},
            discard if not={series}{pegasus-bigram}]
    table [x=evading_reword, y=evading_quality_pass, meta=setting] {plots/data/attack_channels/#1.dat};%
  \dataplot [attackchannels/inline-label, nodes near coords,
            point meta=explicit symbolic,
            every node near coord/.append style={anchor=south west, xshift=4pt, yshift=3pt, text=channelSentence!85!black},
            discard if not={series}{parrot}]
    table [x=evading_reword, y=evading_quality_pass, meta=setting] {plots/data/attack_channels/#1.dat};%
  \dataplot [attackchannels/inline-label, nodes near coords,
            point meta=explicit symbolic,
            every node near coord/.append style={anchor=south west, xshift=1pt, yshift=1pt, text=channelSentence!85!black},
            discard if not={series}{parrot-bigram}]
    table [x=evading_reword, y=evading_quality_pass, meta=setting] {plots/data/attack_channels/#1.dat};%
}

\newcommand{\attackchannellegend}{%
  \addlegendimage{attackchannels/pegasus}
  \addlegendentry{sentence-level controls}%
  \addlegendimage{attackchannels/dipper-order}
  \addlegendentry{\dipper\ order}%
  \addlegendimage{attackchannels/dipper-lexical}
  \addlegendentry{\dipper\ lexical}%
  \addlegendimage{attackchannels/adaptive}
  \addlegendentry{\edaconfigpossent}%
  \addlegendimage{attackchannels/adaptive-bag-sent}
  \addlegendentry{\edaconfigbagsent}%
  \addlegendimage{attackchannels/adaptive-bag-semspans}
  \addlegendentry{\edaconfigbagspan}%
}

%% file: commands.tex
\newcommand{\ebw}{EBW\xspace}
\newcommand{\ebws}{EBWs\xspace}

\newcommand{\semstamp}{SemStamp\xspace}
\newcommand{\ksemstamp}{k-\semstamp}        % k-SemStamp (tracks \semstamp)
\newcommand{\semstampfamily}{(k)-\semstamp}
\newcommand{\pmark}{PMark\xspace}
\newcommand{\samark}{SAMark\xspace}
\newcommand{\unigram}{Unigram\xspace}
\newcommand{\kgw}{KGW\xspace}

\newcommand{\ourstamp}{SwordStamp\xspace}
\newcommand{\papertitle}{Semantic Watermarking with Order-Robust Detection over Sub-sentence Units}

\newcommand{\kourstamp}{k-\ourstamp}        % tracks \ourstamp
\newcommand{\ourstampfamily}{(k)-\ourstamp} % SwordStamp and k-SwordStamp

\DeclareSymbolFont{swordextras}{U}{zavm}{m}{n}
\DeclareMathSymbol{\crossedswords}{\mathord}{swordextras}{124}
\newcommand{\swordmark}{\ensuremath{\crossedswords}}

\newcommand{\semspan}{semspan\xspace}
\newcommand{\semspans}{semspans\xspace}
\newcommand{\Semspan}{Semspan\xspace}
\newcommand{\Semspans}{Semspans\xspace}

\newcommand{\edattack}{embedding displacement attack\xspace}  % our adaptive attack (\Cref{alg:eda})
\newcommand{\Edattack}{Embedding displacement attack\xspace}  % sentence-initial / heading form
\newcommand{\eda}{EDA\xspace}                                  % abbreviation for the embedding displacement attack
\newcommand{\edap}{EDA-P\xspace}  % positional anchors, sentences (\Cref{alg:eda})
\newcommand{\edas}{EDA-S\xspace}  % anchors and segmenter the target publishes (\Cref{alg:edas})
\newcommand{\edad}{EDA-D\xspace}  % detector access: the provider's own detector (\Cref{alg:detector-access})
\newcommand{\edaconfig}[2]{(#1, #2)}
\newcommand{\edaconfigpossent}{\edaconfig{pos}{sent}\xspace}
\newcommand{\edaconfigbagsent}{\edaconfig{bag}{sent}\xspace}
\newcommand{\edaconfigbagspan}{\edaconfig{bag}{span}\xspace}
\newcommand{\bagattack}{bag displacement attack\xspace}
\newcommand{\dipper}{Dipper\xspace}
\newcommand{\pegasus}{Pegasus\xspace}
\newcommand{\parrot}{Parrot\xspace}
\newcommand{\eos}{\textsc{eos}\xspace}      % paraphraser end-of-sequence marker (\Cref{alg:eda})

\newcommand{\llama}{\texttt{Llama-3.1-8B}\xspace}  % provider generator
\newcommand{\qwen}{\texttt{Qwen3-32B}\xspace}      % LLM judge for Q and fidelity

\newcommand{\Setup}{\mathsf{Setup}}
\newcommand{\Watermark}{\mathsf{Watermark}}
\newcommand{\Detect}{\mathsf{Detect}}
\newcommand{\Emb}{\mathrm{Emb}}                 % public sentence encoder
\newcommand{\Embsur}{\widetilde{\mathrm{Emb}}}  % attacker's surrogate encoder
\newcommand{\reg}[2][]{\ensuremath{r_{#1}(#2)}} % region of a sentence: \reg{s}=r(s) for (k-)SemStamp; \reg[t,j]{s}=r_{t,j}(s) for PMark position t and channel j. Glyph isolated here.
\newcommand{\keybit}[2]{\ensuremath{k_{#1,#2}}} % PMark key bit at sentence position #1, channel #2: \keybit{t}{j}=k_{t,j}.
\newcommand{\fkeybit}[1]{\ensuremath{k_{#1}}}   % SAMark key bit, channel #1: drawn per query, constant across the response, so no position index.
\newcommand{\rkeybit}[1]{\ensuremath{\hat{k}_{#1}}} % Key bit that SAMark's self-anchored detector infers from the suspect text by majority vote.

\newcommand{\valfrac}{\gamma}     % valid-region fraction (config: lmbd)
\newcommand{\margin}{\ensuremath{\delta}}   % rejection margin, matches original papers (config: delta)
\newcommand{\qual}{Q}             % content-preservation score, \qual(x,y)
\newcommand{\fid}{F}
\newcommand{\qbar}{\ensuremath{\bar{q}}}    % score cutoff in (\robeps,\qbar)-robustness
\newcommand{\robeps}{\varepsilon} % robustness bound
\newcommand{\budget}{K}           % attacker per-unit search budget
\newcommand{\hits}{H}             % valid-region hit count in z-test
\newcommand{\dep}{d_{\partial}}
\newcommand{\seg}{\mathsf{Seg}}
\newcommand{\spanwin}{w}
\newcommand{\spanmax}{L}
\newcommand{\thresh}{\tau}        % detection threshold function, \thresh(\fpr)
\newcommand{\fpr}{\rho}           % target FPR
\newcommand{\Nprov}{N}            % provider per-sentence candidate budget: rejection maxout cap in (k)-SemStamp, pool size in PMark / best-of-N
\newcommand{\TPR}{\mathrm{TPR}}   % true-positive rate
\newcommand{\AUROC}{\mathrm{AUROC}}
\newcommand{\ASR}{\mathrm{ASR}}
\newcommand{\zret}{\ensuremath{\mathcal{R}}}

\newcommand{\maxnewtok}{256}

\newcommand{\edamaxtok}{512}

\newcommand{\nregions}{8}

\providecommand{\DetectionMetricMode}{tpr1}
\newif\ifDetectionMetricThreshold

\newcommand{\SetDetectionMetric}[1]{%
  \edef\CurrentDetectionMetric{#1}%
  \ifdefstring{\CurrentDetectionMetric}{tpr1}{%
    \DetectionMetricThresholdtrue
    \def\DetectionMetricColumn{TPR1}
    \def\DetectionMetricValueSuffix{TPROne}
    \def\DetectionMetricLabel{\ensuremath{\TPR\mathord{@}1\%\ \mathrm{FPR}}}
    \def\DetectionMetricShortLabel{\ensuremath{\TPR\mathord{@}1\%}}
    \def\DetectionMetricYMin{0}
  }{\ifdefstring{\CurrentDetectionMetric}{tpr5}{%
    \DetectionMetricThresholdtrue
    \def\DetectionMetricColumn{TPR5}
    \def\DetectionMetricValueSuffix{TPRFive}
    \def\DetectionMetricLabel{\ensuremath{\TPR\mathord{@}5\%\ \mathrm{FPR}}}
    \def\DetectionMetricShortLabel{\ensuremath{\TPR\mathord{@}5\%}}
    \def\DetectionMetricYMin{0}
  }{\ifdefstring{\CurrentDetectionMetric}{auroc}{%
    \DetectionMetricThresholdfalse
    \def\DetectionMetricColumn{AUROC}
    \def\DetectionMetricValueSuffix{AUROC}
    \def\DetectionMetricLabel{\ensuremath{\AUROC}}
    \def\DetectionMetricShortLabel{\ensuremath{\AUROC}}
    \def\DetectionMetricYMin{0.5}
  }{%
    \PackageError{semstamp}{Unknown detection metric `#1'}%
      {Use tpr1, tpr5, or auroc in \string\DetectionMetricMode.}
  }}}
}
\SetDetectionMetric{\DetectionMetricMode}

%% file: sections/abstract_arxiv.tex
\begin{abstract}
Semantic watermarks tie the mark to sentence meaning rather than token choices, promising robustness to content-preserving edits.
However, the detector only observes attacker-supplied text, which can be reworded, reordered, or resegmented to evade detection without content loss.
Rewording, reordering, and resegmentation all cause \emph{embedding displacement}: detection tests embeddings different from those selected during watermarking and can therefore lose the mark.
Our adaptive \edattack\ (\eda) admits all three edits under a single objective that maximizes this displacement.
It uses a public paraphraser and surrogate encoder without access to the provider's generator or secret key.
At a \resultPrimaryFpr\ \emph{false-positive rate} (FPR) and \emph{content-preservation threshold} $\qbar=\resultQualityRequirement$, \eda successfully removes the mark on \resultPriorEdapAsrRange\ of documents across four schemes, the highest among the tested attacks.
Therefore, \eda\ evaluates the schemes' robustness more thoroughly than passive paraphrasing.

To address these vulnerabilities, we design \ourstampfamily: \textbf{\underline{s}}emantic \textbf{\underline{w}}atermarks with \textbf{\underline{o}}rder-\textbf{\underline{r}}obust \textbf{\underline{d}}etection over \textbf{\underline{s}}ub-sentence units, reducing sensitivity to attacker-chosen structure at a small quality cost.
Against \kourstamp, the strongest no-box attack we test is an \eda\ variant adapted to its design, with a \resultKSwordEdasAsr\ attack-success rate.
A stronger \eda\ with access to the provider's detector and secret key reaches a \resultWhiteboxKSwordAsrMax\ attack-success rate, compared with \resultWhiteboxKSemAsrMax\ on \ksemstamp.
Our code is available at \url{https://github.com/D-Diaa/SwordStamp}.
\end{abstract}

%% file: sections/intro.tex
\section{Introduction}
\label{sec:intro}

\input{figures/teaser}

A text watermark embeds a hidden signal during generation so that a detector with a secret key can later identify the resulting text as AI-generated.
Providers are beginning to deploy text watermarks.
Google deploys SynthID-Text in Gemini~\cite{dathathri2024synthid}.
Anthropic has announced that future Claude models will use a version of SynthID-Text to meet the European Union's AI Act marking requirements~\cite{anthropic2026textwatermark,europeancommission2026transparency}.

Users may edit generated text before using it in an article, email, or report.
A detector may therefore receive only the edited version.
A useful watermark should remain detectable through meaning-preserving edits while its detector maintains a low FPR on human text.
Evasion alone does not make an attack successful, because an attacker that discards the content evades any detector and gains nothing from the provider.
We score how well an attack output preserves the marked text's content.
$\qbar$ is the minimum content-preservation percentage an attacker must reach.
An attack is \emph{$\qbar$-successful} when its output both evades the detector and scores at least $\qbar$.
Most text watermarks operate through token choices, which paraphrasing can remove at a high $\qbar$~\cite{krishna2023paraphrasing,diaa2024textwatermark}.
Encoder-based watermarks (\ebws) instead mark whole sentences~\cite{hou2024semstamp,hou2024ksemstamp,huo2025pmark,huo2026samark}.
A sentence encoder maps each sentence to a vector, and the secret key selects a set of regions of that embedding space.
\semstampfamily~\cite{hou2024semstamp,hou2024ksemstamp} generate sentences by \emph{rejection sampling}: they draw candidate sentences from the language model and keep the first one whose embedding falls inside a selected region.
The detector re-embeds the sentences of a suspect text and counts how many of them land in selected regions.
The implicit assumption is that meaning-preserving edits tend to keep a sentence's embedding within the same region, so an evading rewrite would require a change in meaning.

Most prior evaluations of \ebws only consider sentence-by-sentence paraphrasing that preserves sentence order and boundaries~\cite{hou2024semstamp,hou2024ksemstamp}.
These evaluations do not cover an attacker who changes sentence structure.
A whole-text attacker can preserve content while reordering sentences, splitting one sentence into two, or merging adjacent sentences.
These edits change either the unit that the detector embeds or the position at which it tests that unit.

We apply rewording, reordering, and resegmentation separately to measure each edit's effect on detection.
Although these edits differ, each causes the detector to evaluate an embedding different from the one selected during watermarking.
Rewording shifts the original sentence's embedding, reordering moves a different sentence into the tested position, and resegmentation changes the unit that the detector embeds.
This shared \emph{embedding displacement} effect is the basis for our \edattack\ (\eda).
\eda observes only the marked text and cannot query the detector or the generator, matching the \emph{no-box} setting of prior work~\cite{diaa2024textwatermark}.
It runs the provider's own procedure with the opposite objective.
The provider samples candidate sentences and keeps one that lands inside a selected region; \eda rewrites the text one sentence at a time and, at each step, keeps the candidate farthest from the source sentence under a public surrogate encoder.
The surrogate differs from every provider encoder we test, yet candidates displaced farther under the surrogate also tend to be displaced farther under the provider's encoder (\Cref{fig:transfer}).
Because each candidate continues a rewrite of the whole text rather than replacing one sentence in isolation, \eda can also reorder content and change sentence boundaries.

We evaluate \eda against four \ebws: \semstamp, \ksemstamp, \pmark, and the concurrent \samark~\cite{hou2024semstamp,hou2024ksemstamp,huo2025pmark,huo2026samark}.
At \resultPrimaryFpr\ FPR and $\qbar=\costasrqualitybar$, \eda is $\qbar$-successful on \resultPriorEdapAsrRange\ of marked documents, the highest success of the attacks we test on each of the four schemes.
\samark\ resists the whole-text paraphraser \dipper~\cite{krishna2023paraphrasing} best among the four, holding it to \resultSamarkDipperAsr, yet \eda\ succeeds on \resultSamarkEdapAsr\ of its documents.
At this operating point, \eda\ reaches a higher ASR on each scheme than every tested watermark-agnostic paraphraser, including \pegasus~\cite{zhang2020pegasus}, \parrot~\cite{damodaran2021parrot}, and \dipper.

To improve the robustness of \ebws, we add three mechanisms to \semstampfamily.
Best-of-$N$ selection scores a fixed pool of candidate sentences and emits the one whose embedding sits farthest inside a selected region, so a rewrite must travel farther to leave it (\S\ref{subsec:best-of-n}).
A fixed valid set uses the same key-selected regions at every position instead of deriving them from the preceding sentence, which makes detection order-robust (\S\ref{subsec:fixed-valid-sets}).
Deterministic sub-sentence units, which we call \emph{\semspans}, follow changes in meaning rather than punctuation and reduce sensitivity to sentence splits and merges (\S\ref{subsec:semspan}).
Applying these mechanisms produces \ourstampfamily.

\Cref{fig:teaser} compares schemes under their strongest tested attacks, including \edas, the \eda variant adapted to every scheme's design (\S\ref{subsec:scheme-attack}).
Across all tested $\qbar$, \kourstamp\ has the lowest attack success among the six schemes.
At \resultPrimaryFpr\ FPR and $\qbar=\resultQualityRequirement$, \ourstamp\ reduces scheme-specific \eda\ attack success by \resultSwordEdasAsrReduction{} percentage points relative to \semstamp, at a fidelity cost of \resultSwordFidelityCost{} percentage points.
At the same operating point, \kourstamp\ reduces scheme-specific \eda\ attack success by \resultKSwordEdasAsrReduction{} percentage points relative to \ksemstamp, at a fidelity cost of \resultKSwordFidelityCost{} percentage points.
\Cref{fig:fidelity-robustness} places this trade-off against all six schemes at the same operating point.

As a stress test beyond the no-box setting, we evaluate detector-access \eda, which receives the provider's detector, including its secret key and encoder.
This attacker evaluates each candidate with the detector and selects the closest candidate that the detector misses.
With four candidate rewrites per step, detector evasion is \resultWhiteboxKSwordEvasionLowBudget\ on \kourstamp, compared with \resultWhiteboxKSemEvasionLowBudget\ on \ksemstamp, showing that evading rewrites are harder to find.
At the largest search budget, detector evasion reaches \resultWhiteboxKSwordEvasionHighBudget, but only \resultWhiteboxKSwordEvadingQualityPassHighBudget\ of them clear $\qbar$ (\Cref{fig:whitebox}).
This caps the attack's $\qbar$-success rate at \resultWhiteboxKSwordAsrMax, compared with \resultWhiteboxKSemAsrMax\ on \ksemstamp, and demonstrates \kourstamp's improved robustness to content-preserving attacks.

We make three contributions:
\begin{enumerate*}[label=(\arabic*), itemjoin={{; }}, itemjoin*={{; and }}]
  \item We measure how rewording, reordering, and resegmentation affect \ebw detection and consolidate them into \eda, a single adaptive whole-text attack guided by embedding displacement
  \item We develop \ourstampfamily\ by extending \semstampfamily\ with best-of-$N$ selection, order-robust detection, and sub-sentence units
  \item We evaluate six watermarks under the $\qbar$-success criterion, measure each mechanism's effects on clean detection, output quality, and attack success, and stress-test \kourstamp\ and \ksemstamp\ under detector access
\end{enumerate*}.

%% file: figures/teaser.tex
\input{figures/fidelity_robustness}

\newcommand{\teaserline}[3]{%
  \IfFileExists{plots/data/teaser/#1.dat}{%
    \dataplot [scheme/#2, error bars/.cd, y dir=both, y explicit]
      table [x=bar, y=y, y error minus=em, y error plus=ep]
      {plots/data/teaser/#1.dat};
    \addlegendentry{#3}%
  }{}%
  \IfFileExists{plots/data/teaser/#1-partial.dat}{%
    \dataplot [scheme/#2, scheme/partial]
      table [x=bar, y=y] {plots/data/teaser/#1-partial.dat};
    \addlegendentry{#3\,$^\dagger$}%
  }{}%
}

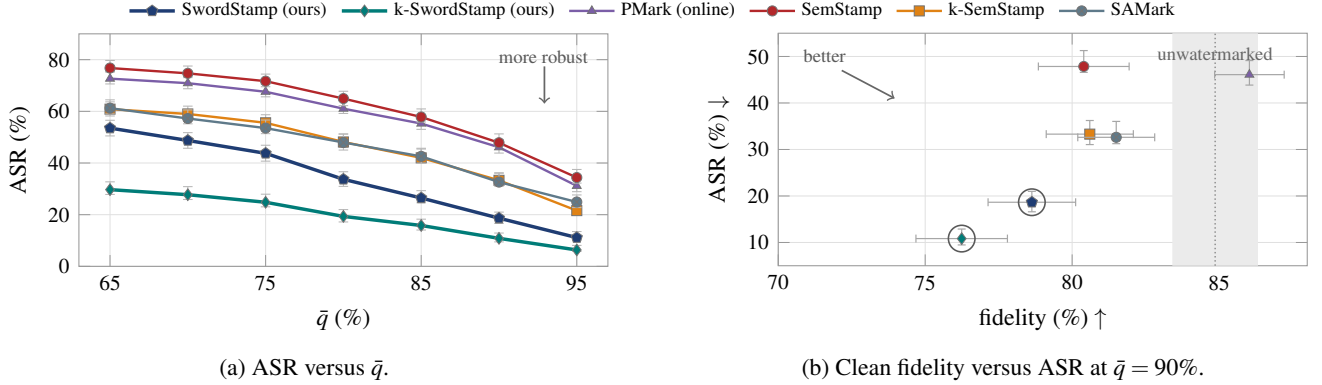
\begin{figure*}[t]
  \centering
  \ref*{legend:teasers}\par
  \begin{subfigure}[t]{0.48\textwidth}
    \centering
    \begin{tikzpicture}
      \begin{axis}[
        width=0.82\linewidth, height=\plotpanelheight,
        scale only axis,
        qbar axis,
        asr axis,
        ymax=0.9,
        error bars/error mark options={rotate=90, mark size=1.8pt,
                                       draw=black!28, line width=0.4pt},
        legend to name=legend:teasers,
        legend columns=6,
        legend style={draw=none, font=\scriptsize, cells={anchor=west},
                      /tikz/every even column/.append style={column sep=3pt}},
      ]
        \teaserline{firmstamp}{firmstamp}{\ourstamp\ (ours)}
        \teaserline{kfirmstamp}{kfirmstamp}{\kourstamp\ (ours)}
        \teaserline{pmark-online}{pmark-online}{\pmark\ (online)}
        \teaserline{semstamp}{semstamp}{\semstamp}
        \teaserline{ksemstamp}{ksemstamp}{\ksemstamp}
        \teaserline{samark}{samark}{\samark}
        \node[anchor=north east, font=\scriptsize, text=black!65]
          at (axis description cs:0.98,0.97) {more robust};
        \draw[->, draw=black!60, line width=0.65pt]
          (axis description cs:0.88,0.86) -- (axis description cs:0.88,0.70);
      \end{axis}
    \end{tikzpicture}
    \caption{ASR versus $\qbar$.}
    \label{fig:teaser-asr}
  \end{subfigure}\hfill
  \begin{subfigure}[t]{0.48\textwidth}
    \centering
    \fidelityasrpanel
    \caption{Clean fidelity versus ASR at $\qbar=\resultQualityRequirement$.}
    \label{fig:fidelity-robustness}
  \end{subfigure}
  \caption{%
    Strongest no-box attack success rate (ASR) at \resultPrimaryFpr\ FPR across attack families and settings.
    Vertical bars give percentile-bootstrap $95\%$ intervals for maximum ASR; horizontal bars in (b) give $95\%$ confidence intervals for clean fidelity.
    The shaded region in (b) gives the unwatermarked fidelity interval, and the dotted line marks its mean.%
  }
  \label{fig:teaser}
\end{figure*}

%% file: figures/fidelity_robustness.tex
\input{plots/data/fidelity_robustness/constants}

\newcommand{\fidrobpoint}[2]{%
  \addplot [scheme/#2, only marks,
            restrict expr to domain={\thisrow{id}}{#1:#1},
            error bars/.cd,
            x dir=both, x explicit,
            y dir=both, y explicit]
    table [x=fidelity, y=asr,
           x error minus=fem, x error plus=fep,
           y error minus=aem, y error plus=aep]
      {plots/data/fidelity_robustness/points.dat};
}

\newcommand{\fidelityasrpanel}{%
  \begin{tikzpicture}
    \begin{axis}[
      paper axis,
      width=0.82\linewidth,
      height=\plotpanelheight,
      scale only axis,
      xmin=70, xmax=88,
      ymin=5, ymax=55,
      xtick={70,75,80,85},
      ytick={10,20,30,40,50},
      xlabel={fidelity (\%) $\uparrow$},
      ylabel={$\ASR$ (\%) $\downarrow$},
      error bars/error bar style={line width=0.45pt, draw=black!35},
      error bars/error mark options={rotate=90, mark size=1.5pt,
                                     draw=black!35, line width=0.45pt},
      clip mode=individual,
    ]
      \addplot [draw=none, fill=black!8, forget plot]
        coordinates {
          (\unwatermarkedfidelitylow,5)
          (\unwatermarkedfidelityhigh,5)
          (\unwatermarkedfidelityhigh,55)
          (\unwatermarkedfidelitylow,55)
        } \closedcycle;
      \draw[draw=black!45, densely dotted, line width=0.55pt]
        (axis cs:\unwatermarkedfidelitymean,5) --
        (axis cs:\unwatermarkedfidelitymean,55);
      \node[anchor=north, font=\scriptsize, text=black!60]
        at (axis cs:\unwatermarkedfidelitymean,54) {unwatermarked};
      \fidrobpoint{0}{semstamp}
      \fidrobpoint{1}{ksemstamp}
      \fidrobpoint{2}{pmark-online}
      \fidrobpoint{3}{samark}
      \fidrobpoint{4}{firmstamp}
      \fidrobpoint{5}{kfirmstamp}
      \addplot [only marks, mark=o, mark size=5pt, forget plot,
                restrict expr to domain={\thisrow{id}}{4:5},
                mark options={draw=black!65, fill=none, line width=0.6pt}]
        table [x=fidelity, y=asr]
          {plots/data/fidelity_robustness/points.dat};
      \node[anchor=north west, font=\scriptsize, text=black!65]
        at (axis description cs:0.03,0.97) {better};
      \draw[->, draw=black!60, line width=0.65pt]
        (axis description cs:0.12,0.84) -- (axis description cs:0.22,0.72);
    \end{axis}
  \end{tikzpicture}
}

%% file: plots/data/fidelity_robustness/constants.tex
\newcommand{\unwatermarkedfidelitymean}{84.866536}
\newcommand{\unwatermarkedfidelitylow}{83.409583}
\newcommand{\unwatermarkedfidelityhigh}{86.323490}
\newcommand{\resultUnwatermarkedFidelity}{84.9\%}
\newcommand{\resultUnwatermarkedFidelityLow}{83.4\%}
\newcommand{\resultUnwatermarkedFidelityHigh}{86.3\%}

%% file: sections/background.tex
\section{Text Watermarking}
\label{sec:background}

We formalize two types of text watermarking: token-level watermarks and encoder-based watermarks.
\Cref{tab:notation} summarizes our notation.

\textbf{Watermarking and detection.}
An autoregressive language model $\mathcal{M}$ defines a distribution $P_{\mathcal{M}}(x_t\mid x_{<t})$ over the next token and generates text by sampling from it left to right.
We define a \emph{text watermark} as a triple of randomized algorithms $(\Setup, \Watermark, \Detect)$ over a key space $\mathcal{K}$ and a text space $\mathcal{Y}$.
$\Setup \to k \in \mathcal{K}$ draws a secret key.
$\Watermark(k) \to x \in \mathcal{Y}$ samples text from $\mathcal{M}$ so that $k$ leaves a statistical trace in $x$.
$\Detect(k, y) \to z \in \mathbb{R}$ maps a suspect text $y$ and key $k$ to a score $z$, where a larger $z$ indicates a stronger trace of $k$.
We write $z_k(y)$ for the score that $\Detect$ assigns to text $y$ under key $k$.
The target FPR $\fpr$ is the probability that the detector flags unwatermarked text.
The provider calibrates the corresponding score threshold $\thresh(\fpr)$, and the detector flags $y$ as watermarked when $z_k(y)>\thresh(\fpr)$.
Concretely, the \emph{null distribution} is the distribution of $z_k$ on unwatermarked text, and $\thresh(\fpr)$ is its $(1-\fpr)$-quantile.
The provider estimates this null from a held-out sample of human-written text, or computes it analytically when it has a closed form.
We study \emph{zero-bit} watermarks, which embed one signal, so $\Detect$ reports only the key's trace, not an embedded message.

\textbf{Token-level watermarks.}
Token-level watermarks operate on the model's vocabulary $V$, treating the output as a token sequence.
The green-red scheme of Kirchenbauer et al.~\cite{kirchenbauer2023watermark} (\kgw) hashes a window of preceding tokens to seed a pseudo-random split of $V$ at each position.
Each split partitions $V$ into a \emph{green list} of size $\valfrac\lvert V\rvert$ and a red complement.
\kgw\ then adds a fixed logit bias $\delta$ to green tokens so the model emits them more often.
Throughout, $\valfrac\in(0,1)$ is the fraction of the space that the key marks \emph{valid} at one position.
Detection applies the following one-proportion $z$-test to the green-token count $\hits$ among $n$ tokens.
\begin{equation}
  z = \frac{\hits - \valfrac n}{\sqrt{n\,\valfrac(1-\valfrac)}}
  \label{eq:ztest}
\end{equation}
The statistic standardizes the green-token count, so the unwatermarked null centers at zero.
We call such a partition \emph{context-dependent}: the green-red split is re-derived at every position from the preceding tokens.
Substituting a token can turn a green-token hit into a miss at that position.
The substitution also re-seeds subsequent splits, so one edit can disrupt the mark at later positions.
The \unigram scheme of Zhao et al.~\cite{zhao2024provable} instead uses a \emph{fixed} split, one green list for the entire generation, independent of context.
Zhao et al.\ prove that under the fixed split each token edit changes the detection score by a bounded amount.
For sufficiently high-entropy text, evading detection therefore requires a number of edits that grows linearly with the text length.
Meaning-preserving rewording can nevertheless change enough token choices to evade token-level watermarks~\cite{krishna2023paraphrasing,diaa2024textwatermark}.

\input{tables/notation}

\textbf{Encoder-based watermarks.}
A second family responds to this fragility by raising the unit of marking from the token to the sentence.
These schemes treat text as a sentence sequence $x=(s_1,\dots,s_n)$ and assign the mark to sentence meaning~\cite{hou2024semstamp,hou2024ksemstamp,huo2025pmark,huo2026samark}.
An encoder $\Emb$ maps each sentence $s$ to a vector $\Emb(s)\in\mathbb{R}^h$, and a partition $P=\{R_1,\dots,R_c\}$ splits $\mathbb{R}^h$ into $c$ \emph{regions}, so each sentence inherits the region of its embedding.
We write $\reg{s}$ for the \emph{region of $s$}.
Every scheme we study depends on a sentence encoder to place and verify its mark, so we call the family \emph{encoder-based watermarks} (\ebws).
The detector operates only on the suspect text and tests whatever sentence embeddings it recovers.
Under a fixed partition $P$, a recovered sentence $s'$ that rewrites $s$ in-place keeps $\reg{s'}=\reg{s}$ whenever it moves the embedding less than the distance from $\Emb(s)$ to the nearest boundary.
The \emph{\ebw robustness premise} is that a content-preserving rewrite usually moves the embedding by less than that distance, so the mark survives rewording.

\semstamp~\cite{hou2024semstamp} forms the partition with locality-sensitive hashing (LSH)~\cite{indyk1998approximate,charikar2002similarity}: $b$ random hyperplanes through the origin induce $c=2^b$ regions.
At step $t$, the previous sentence's region $\reg{s_{t-1}}$ seeds a pseudo-random split of the region indices $\{1,\dots,c\}$ into a \emph{valid} set $G_t$ of size $\valfrac c$ and a \emph{blocked} set $B_t$, the semantic counterpart of green and red.
This seeding makes the partition context-dependent in the same sense as \kgw, with $\reg{s_{t-1}}$ in place of the preceding-window hash.
The provider generates $s_t$ by rejection sampling, drawing candidates from $\mathcal{M}$ and accepting the first whose region is valid, $\reg{s_t}\in G_t$.
A margin $\margin$ additionally rejects any candidate whose embedding lies within $\margin$ of a region boundary.
\semstamp\ caps rejection at a \emph{maxout} budget of $\Nprov$ candidates, then emits the last draw even when it fails the valid-region or margin condition.
\ksemstamp~\cite{hou2024ksemstamp} replaces the hyperplanes with $k$-means clusters: (i) the $c$ regions are the Voronoi cells of centroids fitted to a domain corpus, (ii) $\reg{s}$ is the index of the centroid nearest $\Emb(s)$, and (iii) the margin becomes the minimum gap to the next-nearest centroid.

Following \semstamp's private-encoder assumption~\cite{hou2024semstamp}, our threat model treats the deployed encoder checkpoint $\Emb$, partition geometry, and large prime $p$ as provider-private key material for both \semstamp\ and \ksemstamp.
\semstamp\ draws its hyperplanes at $\Setup$, whereas \ksemstamp\ fits its centroids to a domain corpus and keeps them private.
Both schemes fine-tune $\Emb$ contrastively on paraphrase pairs to reinforce the robustness premise.
To verify a suspect text, both schemes segment it into sentences and re-derive each step's valid set from the suspect text itself, seeding $G_t$ on the recovered $\reg{s_{t-1}}$.
The detector then counts the valid-region hits $\hits$ among the $n$ recovered sentences and applies the same one-proportion $z$-test of \Cref{eq:ztest}.
We call \semstamp\ and \ksemstamp\ the \emph{\semstampfamily family}.

\pmark~\cite{huo2025pmark} computes its partition geometry from candidates sampled from $\mathcal{M}$ at each sentence position rather than fixing it at $\Setup$.
A \emph{channel} is a secret vector $v_j\in\mathbb{R}^h$ with \emph{proxy function} $f_j(s) = \langle v_j,\Emb(s)\rangle$; \pmark\ maintains $b$ mutually orthogonal channels.
At each position, \pmark\ samples $\Nprov$ candidates from $\mathcal{M}$ and processes them through the $b$ channels in sequence.
In channel $j$, it sets the position-specific boundary $m_{t,j}$ to the median of $f_j(s)$ over candidates that survived the preceding channels.
We write $\reg[t,j]{s}\in\{0,1\}$ for the \emph{region of $s$ at position $t$ in channel $j$}, the side of the position-specific boundary $m_{t,j}$ on which $f_j(s)$ falls.
We write $\keybit{t}{j}\in\{0,1\}$ for the \emph{key bit} that designates the valid side, drawn independently and uniformly at random for each position $t$ and channel $j$.
\pmark\ then retains the half whose region matches $\keybit{t}{j}$.
Sampling uniformly from the final survivors leaves the model's output distribution unchanged in expectation over the keys, so \pmark\ is \emph{distortion-free} by construction~\cite{huo2025pmark}.
Since each median splits its input in half, a sentence drawn from $\mathcal{M}$ falls on the valid side of a channel with probability $1/2$.

Detection re-estimates these per-position medians from $\mathcal{M}$: at each position $t$, \pmark\ resamples $\Nprov$ candidates conditioned on the preceding context and takes $m_{t,j}$ as the median of $f_j$ over that pool.
Sentence $s_t$ contributes channel evidence according to $\reg[t,j]{s_t}$ and $\keybit{t}{j}$.
\pmark\ aggregates the $bn$ channel-sentence evidence terms with a soft $z$-test version of \Cref{eq:ztest}.
\pmark\ uses a standard-normal approximation to set $\thresh(\fpr)$ rather than estimating it from human text.
Note that detection requires querying $\mathcal{M}$ to re-estimate the medians.
\pmark\ also defines an \emph{offline} variant that uses a fixed prior median score as a boundary, removing the need for generator queries at detection time but giving up the distortion-free guarantee.
Huo et al.\ report lower robustness for the offline variant under paraphrase attack~\cite{huo2025pmark}.
Because robustness to paraphrase attack is our central question, we study the more robust online variant throughout.

Concurrent work, \samark~\cite{huo2026samark}, extends the offline variant to improve robustness.
\samark\ replaces \pmark's per-position key bit $\keybit{t}{j}$ with one bit $\fkeybit{j}$ per channel, drawn once per query and held constant across the response.
Every sentence then scores against the same bit, so the detector is order-robust, matching the property that \unigram\ provides at the token level.
Unlike \unigram's green list, however, these bits are not long-lived key material: \samark\ never shares them with the detector.
Detection is therefore \emph{self-anchored}: the detector infers each bit by majority vote over the suspect text, setting $\rkeybit{j}=1$ exactly when $\sum_{i=1}^{n} f_j(s_i)>0$.
The vote returns whichever side the text already favors, so the score rewards any text whose sentences concentrate on one side of a channel, watermarked or not.
\samark\ estimates $\thresh(\fpr)$ from a human-text sample.

%% file: tables/notation.tex
\begin{table}[t]
  \centering
  \caption{Notation used throughout the paper.}
  \label{tab:notation}
  \small
  \begin{tabular}{@{}lp{0.72\columnwidth}@{}}
    \toprule
    Symbol & Meaning \\
    \midrule
    $x$, $y$ & marked text; attacked text $\mathcal{A}(x)$ \\
    $k$ & secret watermarking key \\
    $\Emb$, $\Embsur$ & provider encoder; attacker surrogate \\
    $P$, $\reg{s}$ & partition into $c$ regions; region index of $s$ \\
    $\seg$ & segmenter into sentence or \semspan units \\
    $G$, $B$ & valid and blocked region sets; $G_t$ per step \\
    $\valfrac$ & valid-region fraction; null hit rate \\
    $\Nprov$, $\budget$ & provider and attacker candidate budgets \\
    $n$, $\hits$ & scored units; valid-region hits \\
    $z$, $\fpr$, $\thresh(\fpr)$ & detection score; target FPR; calibrated threshold \\
    $\qual(x,y)$, $\qbar$ & content-preservation percentage; minimum threshold \\
    $\fid(x)$ & fidelity percentage of marked text, no reference \\
    \bottomrule
  \end{tabular}
\end{table}

%% file: sections/problem.tex
\section{Threat Model}
\label{sec:problem}

We frame our evaluation as a game between a provider that watermarks generated text and an attacker that seeks to remove the watermark while preserving its content.
We specify each party's knowledge, capabilities, and success condition.

\textbf{Parties.}
The \emph{provider} controls a language model that generates text in response to user prompts.
At generation time, the provider runs the watermarking algorithm with a secret \emph{key} $k$.
At verification time, the provider scores a suspect text with $\Detect$ and flags it when the score exceeds the FPR-calibrated threshold $\thresh(\fpr)$ of \S\ref{sec:background}.
Following Kerckhoffs's principle, the algorithm itself is public.
The \emph{attacker} receives marked text $x$ from the provider, applies a perturbation $\mathcal{A}$ to produce attacked text $y=\mathcal{A}(x)$, and seeks to drive the detector score to or below $\thresh(\fpr)$ while preserving the content of $x$.
We study watermark removal only.
Forging a mark onto text that the provider did not generate and extracting the secret key are separate adversarial objectives~\cite{zhao2025sok} that lie outside our scope.

\textbf{Attacker capabilities.}
Our primary attacker is \emph{no-box}~\cite{diaa2024textwatermark}: it observes only the released text, does not obtain a corpus of marked texts across queries, and cannot query the detector or generator.
A no-box attacker may run its own models, including open paraphrasers and a sentence encoder.
We restrict the attacker to lower-fidelity generators because an equally capable generator could produce the desired unwatermarked text instead of scrubbing the provider's output.
Attackers also differ in whether they exploit knowledge of the watermarking algorithm.
A \emph{passive} attacker applies watermark-agnostic, content-preserving perturbations, whereas an \emph{adaptive} attacker optimizes its perturbation against the known algorithm, following the adaptive-attack paradigm~\cite{diaa2024textwatermark, lukas2024imagewatermark}.
Every attacker is \emph{effort-constrained}: depending on the attack family, its budget counts model calls, candidate perturbations, or optimization steps.
Each attack we instantiate has its own effort parameter, and we trace how attack success changes as the budget grows.
A detector-access stress test grants the attacker access to the provider's secret key (\S\ref{subsec:eval-informed}).

\textbf{Attack success.}
Attack success requires both detector evasion and content preservation; discarding content can trivially evade detection but fails the quality constraint~\cite{zhao2025sok}.
The content-preservation percentage score $\qual(x,y)\in[0\%,100\%]$ measures how well attacked text $y$ preserves marked text $x$.
At target FPR $\fpr$, a $\qbar$-successful attack lowers the detection score to or below $\thresh(\fpr)$ while retaining $\qual(x,y)\ge\qbar$.
For a perturbation strategy $\mathcal{A}$, this joint event has probability
\begin{equation}
  p_{\mathcal{A}}(\fpr,\qbar)
  = \Pr_{\substack{k\sim\Setup\\x\sim\Watermark(k)\\y=\mathcal{A}(x)}}\!\left[
    z_k(y)\le\thresh(\fpr) \wedge \qual(x,y)\ge\qbar
  \right].
  \label{eq:robustness}
\end{equation}
The probability is over the secret key $k$, marked text $x$ generated by $\Watermark(k)$ on prompts drawn from a fixed distribution, the randomness of $\mathcal{A}$, and, for \pmark, the detector's resampling.
A watermark is ($\robeps$,~$\qbar$)-\emph{robust} against $\mathcal{A}$ at FPR $\fpr$ when $p_{\mathcal{A}}(\fpr,\qbar)\le\robeps$.
We estimate $p_{\mathcal{A}}(\fpr,\qbar)$ with the \emph{attack success rate} ($ASR$@$\fpr$), the percentage of marked texts on which the attack is $\qbar$-successful at that FPR\@.
Each attack $\mathcal{A}$ runs at a fixed effort budget, so $p_{\mathcal{A}}(\fpr,\qbar)$ characterizes that attack at that budget alone.
Robustness is therefore a frontier over attack families, effort budgets, and $\qbar$ values.
We report ASR versus $\qbar$ throughout (\S\ref{sec:eval}).

\textbf{Provider success.}
The provider seeks to preserve output fidelity before attack while minimizing ASR across the attack frontier.
\emph{Fidelity} $\fid(x)\in[0\%,100\%]$ is the quality of marked text $x$ judged on its own, with no reference text.
$\fid$ assigns a percentage score to one text, whereas $\qual$ scores an attacked text against the marked text it came from.
\S\ref{subsec:eval-setup} describes how we measure both.

%% file: sections/attacks.tex
\section{Edit Sensitivity and \eda}
\label{sec:attacks}

The \ebw robustness premise covers in-place \emph{rewording}, not two structural changes available to an attacker who controls the whole text: \emph{reordering} changes the context or position that determines a test, and \emph{resegmentation} changes the units the detector recovers.
We first use controlled edits to measure each edit's effect, then consolidate them into \eda, a general adaptive no-box attack guided by \emph{embedding displacement}.
\textbf{Reordering.}
An attacker can reorder a marked text's sentences and adjust its references and connectives while preserving content (\S\ref{sec:problem}).
For context-dependent watermarks, reordering changes the predecessor that keys region tests even when it leaves every sentence unchanged.
Let $\sigma$ map each destination position $t$ to the source index $\sigma(t)$, so the reordered sequence is $(s_{\sigma(1)},\dots,s_{\sigma(n)})$.
For $t>1$ with $\sigma(t)>1$, generation selected $s_{\sigma(t)}$ under the valid set keyed by its original predecessor region $\reg{s_{\sigma(t)-1}}$.
The detector instead tests it under the set keyed by its reordered predecessor region $\reg{s_{\sigma(t-1)}}$.
If these predecessor regions differ, the detector uses a pseudo-randomly unrelated valid set, so $s_{\sigma(t)}$ passes with probability $\valfrac$ over the secret split.
If $\sigma(t)=1$, the detector newly scores the original seed; a sentence moved to the first position instead becomes the unscored seed.
When reordering changes the predecessor region across the scored transitions, the expected hit rate returns to $\valfrac$ and the expected $z$-score approaches zero.

For \pmark, the channel test at position $t$ compares $\reg[t,j]{s_{\sigma(t)}}$ against the position-keyed bit $\keybit{t}{j}$.
Because $\keybit{t}{j}$ is independent and uniform, a moved sentence matches with probability $1/2$.
A permutation that moves every sentence therefore returns the hit count to random, again driving the expected $z$-score toward zero.
\pmark's dynamic partition adds a second reordering-induced mismatch (\S\ref{sec:background}).
At position $t$, the detector re-estimates each median from continuations conditioned on the reordered prefix.
Even a sentence that remains in place can therefore be scored against a median different from the one used during marking.
This suggests that \pmark's robustness to in-place rewording may not transfer to whole-text transformations.
\Cref{subsec:eval-robustness} tests this prediction.
Token watermarks are less exposed to reordering: adjacent tokens are rarely independent, and reordering them typically does not preserve meaning.

\textbf{Resegmentation.}
Resegmentation changes the units that the detector embeds because the detector segments the suspect text itself (\S\ref{sec:background}).
Merging $s_t$ with $s_{t+1}$ produces a unit $m$ whose region $\reg{m}$ need not match $\reg{s_t}$ or $\reg{s_{t+1}}$.
Splitting $s_t$ into fragments $f_1$ and $f_2$ produces regions $\reg{f_1}$ and $\reg{f_2}$, neither of which needs to match $\reg{s_t}$.
Under a context-dependent scheme, a split disturbs three region tests.
The detector evaluates $f_1$ and $f_2$ against the valid sets at their new positions, then tests the following sentence against a set keyed to $\reg{f_2}$ rather than $\reg{s_t}$.
A merge can disturb two region tests.
The detector tests the merged unit, then keys the following sentence's valid set to $\reg{m}$ rather than $\reg{s_{t+1}}$.
Merging also lowers the unit count $n$ that scales the $z$-score of \Cref{eq:ztest}.
Token watermarks have no sentence-like units to split or merge.

\textbf{Controlled edits.}
We isolate each edit type's effect on the detection score with a separate controlled edit.
Each controlled edit changes one of wording, sentence order, or sentence boundaries while inducing negligible change in the other two.
\emph{Reorder} permutes sentence positions and leaves every sentence's text unchanged.
\emph{Split} cuts sentences at an interior point, and \emph{merge} joins adjacent pairs, respectively increasing and decreasing the unit count $n$.
We evaluate both with one direct resegmentation measure.
\emph{Synonym substitution} replaces words in place, leaving sentence order and boundaries intact, and supplies the rewording control.
We report only detection for these controlled edits, since they are not usable attacks (\S\ref{sec:eval}).

\textbf{Embedding displacement.}
\emph{Embedding displacement} is the distance between the embedding selected for a detector position during marking and the embedding tested at that position after editing.
We use greater displacement as a no-box surrogate for a greater chance of crossing a watermark-region boundary.
Rewording shifts the unit's embedding while retaining its position.
Reordering moves a different unit's embedding into the position.
Resegmentation redistributes the text across detector units, replacing the embeddings attached to one or more positions.
Reordering and resegmentation can also change the preceding context that determines the valid set or channel boundary.
Without using the provider's encoder or secret key, our no-box attacker maximizes this distance under a surrogate encoder.
The attack assumes that displacement under $\Embsur$ transfers to displacement under the provider's encoder.
\Cref{fig:transfer} tests this assumption for \semstamp\ and \pmark, using a surrogate encoder with a different architecture from both providers.
\input{figures/transfer}

\textbf{\Edattack.}
\Cref{alg:eda} gives \edap.
It reverses the watermark's sampling objective: the provider searches $\Nprov$ sampled candidates for a sentence in a valid region (\S\ref{sec:background}), whereas the attacker searches $\budget$ paraphrases for the candidate farthest from the source sentence at that position.
We use cosine distance $d(u,v)=1-\langle u,v\rangle/(\lVert u\rVert\,\lVert v\rVert)$.

\begin{algorithm}[t]
\caption{Positional \eda\ (\edap).}
\label{alg:eda}
\begin{algorithmic}[1]
\Require marked text $x=(s_1,\dots,s_n)$; paraphraser $\pi$; surrogate encoder $\Embsur$; budget $\budget$
\Ensure attacked text $y=(s'_1,\dots,s'_{n'})$
\For{$t = 1, 2, \dots$ \textbf{until} $\pi$ emits \eos\ or $y$ reaches \edamaxtok\ tokens} \label{line:loop}
  \State $\mathcal{C} = \{c_1,\dots,c_{\budget}\} \ni c_i \sim \pi(\,\cdot \mid x,\, s'_1,\dots,s'_{t-1},\, t)$ \label{line:sample}
  \State $s'_t \gets \operatorname*{arg\,max}_{c \in \mathcal{C}} d\!\big(\Embsur(c), \Embsur(s_{\min(t,n)})\big)$ \label{line:select}
\EndFor
\State \Return $y = (s'_1,\dots,s'_{n'})$
\end{algorithmic}
\end{algorithm}

The attack generates output positions $t=1,2,\dots$ until the paraphraser emits the end-of-sequence marker \eos\ or the rewrite reaches \edamaxtok\ tokens (line~\ref{line:loop}), allowing the rewrite to differ in length from $x$.
At each position, the attack samples $\budget$ candidate rewrites from an instruction-tuned paraphraser $\pi$ (line~\ref{line:sample}).
The paraphraser conditions on the full marked text $x$ and the previously generated rewrites $s'_1,\dots,s'_{t-1}$.
The \emph{anchor} is source sentence $s_t$ when $t \le n$ and the final source sentence $s_n$ otherwise.
The \emph{surrogate encoder} $\Embsur$ substitutes for the provider's encoder.
The attack selects the candidate farthest from its anchor under $\Embsur$ (line~\ref{line:select}).
Anchoring output position $t$ on source sentence $s_t$ targets a detector that keys each position separately, as \semstampfamily schemes and \pmark\ do (\S\ref{sec:background}).
We call this positional configuration \edap\ when comparing anchor rules.

Conditional provider-encoder displacement rises with surrogate displacement for both schemes.
This pattern supports surrogate selection for \semstamp's contrastively tuned encoder and \pmark's off-the-shelf encoder (\S\ref{sec:background}).
At position $t$, a candidate conditioned on the full marked text and the rewrite so far need not reword $s_t$ in place.
The candidate can split $s_t$, merge it with the following sentence, or advance content from a later sentence and defer the remainder.
The displacement objective admits these structural changes without a separate edit operation.
One attack can therefore combine them with the in-place rewording that every candidate applies.

\edap\ requires no key, partition geometry (hyperplanes, centroids, or \pmark\ channels), or provider encoder, so it runs against every \ebw we evaluate.

%% file: figures/transfer.tex
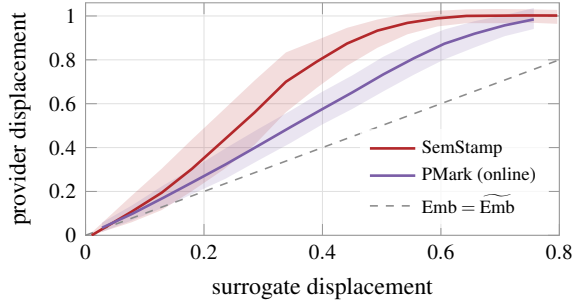
\begin{figure}[t]
  \centering
  \begin{tikzpicture}
    \begin{axis}[transfer/axis]
      \addplot [transfer/identity, forget plot, domain=0:0.8] {x};
      \transferband{semstamp}{schSemStamp}{\semstamp}
      \transferband{pmark}{schPMark}{\pmark\ (online)}
      \addlegendimage{transfer/identity}
      \addlegendentry{$\Emb = \Embsur$}
    \end{axis}
  \end{tikzpicture}
  \caption{%
    Provider-encoder displacement conditioned on surrogate displacement for sentence pairs from $\resultTransferDocs$ documents per scheme.
    Lines mark medians; bands span the middle $50\%$.%
  }
  \label{fig:transfer}
\end{figure}

%% file: sections/method.tex
\section{Robustness Mechanisms for \ebws}
\label{sec:design}
The attacks of \S\ref{sec:attacks} expose three choices that determine a mark's sensitivity to whole-text edits: how far the selected candidate lies inside a valid region, how the valid set changes with context, and which units the provider marks and the detector scores.
Our design changes all three: best-of-$N$ prefers candidates farther inside a valid region, a fixed valid set removes context dependence, and sub-sentence semantic units called \semspans replace sentences.
Best-of-$N$ and the fixed valid set leave $P$ unchanged, while \semspan configurations partition \semspan embeddings rather than sentence embeddings.
Because each modification operates only through region indices and boundary-margin scores, the full design applies to \semstamp's hyperplanes and \ksemstamp's centroids, yielding \ourstamp\ and \kourstamp, respectively.

Throughout, the provider and detector share a deterministic segmenter $\seg$, and $\seg(y)=(u_1,\dots,u_n)$ denotes the units it recovers from text $y$.
The \semstampfamily schemes use $\seg$ to recover sentences.
Because every modification is public (\S\ref{sec:problem}), we state the attacker's response beside each construction and combine those responses into the \edas of \S\ref{subsec:scheme-attack}.

\subsection{Best-of-\texorpdfstring{$N$}{N} Selection}
\label{subsec:best-of-n}

An in-place rewording removes a hit only by crossing a region boundary (\S\ref{sec:attacks}).
The \emph{signed boundary depth} $\dep(u)$ is the watermark's boundary-margin score for unit $u$.
It is positive when $\reg{u}$ lies in the active valid set and negative when $\reg{u}$ lies in the blocked set.
Its magnitude is the minimum absolute hyperplane similarity for LSH and the gap between the nearest and next-nearest centroid distances for $k$-means.
Rejection sampling accepts candidates with $\dep(u)\ge\margin$ under the rule in \S\ref{sec:background}.
Without the key, an attacker cannot target $G$ or $B$, so a content-preserving rewrite may move its embedding across a boundary in either direction (\Cref{fig:best-of-n}).
A deep valid selection gives the rewrite more room before leaving $G$; a shallow blocked fallback gives the rewrite a chance to enter $G$.
\input{figures/best_of_n}

\textbf{Construction.}
Best-of-$N$ evaluates the full pool $\mathcal{C}_t$ of $\Nprov$ candidates and emits its deepest member.
\begin{equation}
  u_t = \operatorname*{arg\,max}_{u\in\mathcal{C}_t} \dep(u)
  \label{eq:best-of-n}
\end{equation}
By ranking every valid candidate above every blocked candidate, it returns the deepest valid candidate when $\mathcal{C}_t$ contains one, and otherwise the blocked candidate nearest a boundary.
The ranking replaces the margin test and maxout's last-draw fallback.
If $\mathcal{C}_t$ contains a candidate that rejection sampling would accept, best-of-$N$ emits a valid unit at least as deep as that candidate.
In that case, best-of-$N$ weakly dominates rejection sampling's margin criterion on the same pool.

Best-of-$N$ trades this depth for provider draws.
If each draw is independently valid with probability $\valfrac$ and we ignore the margin and cap, rejection sampling evaluates $1/\valfrac$ candidates in expectation.
Best-of-$N$ evaluates $\Nprov$ candidates, or $\valfrac\Nprov$ times that expectation.
The two spend those draws differently: rejection sampling issues them one at a time and stops at the first acceptance, whereas best-of-$N$ fixes the pool in advance and can issue and score it in one batch.
Selecting an extreme-depth candidate may affect fidelity; \S\ref{subsec:eval-costs} quantifies that effect.

\textbf{Attacker response.}
Best-of-$N$ exposes no new signal in the marked text, so it requires no design-specific change to \eda beyond sweeping the paraphrase budget $\budget$ (\S\ref{subsec:eval-setup}).

\subsection{Fixed Valid Sets}
\label{subsec:fixed-valid-sets}

Context-dependent valid sets have the reordering weakness in \S\ref{sec:attacks}; a fixed set removes predecessor dependence.

\textbf{Construction.}
A \emph{fixed valid set} is one key-dependent subset of regions that replaces every per-position $G_t$.
It transfers \unigram's context-independent split~\cite{zhao2024provable} from vocabulary items to regions.
At $\Setup$, the provider derives $G\subseteq\{1,\dots,c\}$ from $k$ with $\lvert G\rvert=\valfrac c$, sets $B=\{1,\dots,c\}\setminus G$, and uses $G$ at every generation and detection position.
The detector then counts recovered units in $G$.
\begin{equation}
  \hits_G(y) = \sum_{u\in\seg(y)} \mathbf{1}\!\left[\reg{u}\in G\right]
  \label{eq:fixed-hit-count}
\end{equation}
The detector inserts $\hits_G(y)$ and $n=\lvert\seg(y)\rvert$ into \Cref{eq:ztest} and calibrates $\thresh(\fpr)$ on unmarked text as in \semstampfamily.

Because \Cref{eq:fixed-hit-count} does not depend on unit position, a permutation changes neither $\hits_G(y)$ nor $n$, and moving one unit cannot re-key another test.
An in-place rewrite loses a hit only when its embedding crosses a region boundary into $B$.
Changing unit boundaries avoids that condition: a merged or split unit need not retain the original units' valid-set membership, and merging reduces $n$ in \Cref{eq:ztest}.

\textbf{Null inflation.}
A fixed valid set can also inflate an unmarked passage's score.
Topically concentrated passages place many units in few regions, so a passage whose occupied regions fall disproportionately in $G$ can exceed the null hit rate $\valfrac$ without carrying a watermark.
Calibrating $\thresh(\fpr)$ on unmarked text preserves the target FPR $\fpr$, but can lower detection on unattacked marked text.
\S\ref{subsec:eval-costs} reports the calibrated threshold for every configuration.
By contrast, \samark's self-anchored detector infers key bits from the suspect text (\S\ref{sec:background}), so any concentrated passage inflates its null.

\textbf{Diversity ranking.}
Using one $G$ at every position lets consecutive valid units share a region, risking repetition.
The \emph{diversity ranking} classifies a candidate as green when its region lies in $G$ and differs from the predecessor's region, yellow when it shares the predecessor's valid region, and red when its region lies in $B$.
Writing $\mathrm{color}(u)\in\{2,1,0\}$ for green, yellow, and red, best-of-$N$ lexicographically maximizes $(\mathrm{color}(u),\dep(u))$.
Color decides first, and depth breaks ties within a color.
The diversity ranking does not affect detection: yellow and green units are counted as hits, preserving the order robustness of \Cref{eq:fixed-hit-count}.

\textbf{Attacker response.}
\edap measures distance from the positional anchor $s_t$ (\S\ref{sec:attacks}), but an order-robust detector uses no position in its hit count.
The \bagattack\ scores candidate $u$ against all source units in $x$.
\begin{equation}
  \Delta_{\mathrm{bag}}(u,x)
  = \min_{a\in\seg(x)} d\!\left(\Embsur(u),\Embsur(a)\right)
  \label{eq:bag}
\end{equation}
The minimum grows only when $u$ is far from every source unit, so copying a source unit at another position earns no credit.
At each output step, the attacker emits the candidate that maximizes $\Delta_{\mathrm{bag}}(u,x)$, retaining the budget $\budget$, surrogate encoder $\Embsur$, and whole-text generation of \Cref{alg:eda}.
The attacker can raise the minimum by moving a unit in place, which trades against $\qbar$, or by merging or splitting source units.

\textbf{Secrecy of the valid set.}
Fixing $G$ reintroduces the fixed-list structure that Hou et al.~\cite{hou2024semstamp} identify as \unigram's weakness.
However, a no-box attacker cannot recover $G$.
Doing so requires a corpus of marked texts collected across many provider queries, a black-box capability that the threat model excludes (\S\ref{sec:problem}).
An attacker that pays this cost can also steal and scrub a context-dependent partitioning, as demonstrated for token watermarks~\cite{jovanovic2024watermark,pang2024no}.
Costs differ by a constant factor: a fixed valid set is a histogram over $c$ regions, whereas a context-dependent rule is a $c\times c$ transition matrix.
At $c=\nregions$, this constant-factor difference does not create a distinct security boundary~\cite{hou2024semstamp,hou2024ksemstamp}.

Additionally, a recovered token green list identifies individual tokens that the attacker can replace.
A recovered $G$ instead reveals the valid embedding regions, and its complement $B$ reveals the blocked regions.
To remove a hit on unit $u$, the attacker must either discard $u$, rewrite it into $B$, or resegment its content into units in $B$.
Discarding $u$ removes content.
An in-place rewrite challenges the \ebw robustness premise: it must cross a region boundary while the resulting passage $y$ still satisfies $\qual(x,y)\ge\qbar$.
Thus, $B$ identifies the target region but does not produce a rewrite that reaches it while preserving content.
The next subsection addresses resegmentation.

\subsection{\Semspans}
\label{subsec:semspan}

Fixed valid sets make detection order-robust, but they do not remove segmentation sensitivity: $\hits_G(y)$ stays unchanged only while $\seg(y)$ recovers the same units.
Sentence splitting and merging change those units without necessarily changing passage content (\S\ref{sec:attacks}).
We therefore mark \semspans that an encoder-guided segmenter delimits.
\Cref{fig:semspan} illustrates the recursive cut rule on one sentence.
\input{figures/semspan}

\textbf{Construction.}
A \semspan is a contiguous word sequence within one sentence, delimited at a cut where the encoder observes the largest local embedding change.
The segmenter $\seg$ splits a text into sentences, then recursively cuts each sentence longer than $\spanmax$ words.
A cut point is \emph{admissible} when at least $\spanwin$ words lie on either side and the word to its left is not a function word.
The window condition supplies $\spanwin$ words for each side of the embedding comparison.
The function-word condition keeps a determiner or preposition with the phrase that follows it.
At each admissible point, $\seg$ embeds the $\spanwin$ words on either side and scores their distance.
It selects the highest-scoring point and recurses on both sides until no \semspan exceeds $\spanmax$.
In the example, the first cut lands at the clause boundary, and recursion re-cuts only the resulting \semspan that still exceeds $\spanmax$.

This segmenter has three effects.
First, finer units increase the unit count $n$.
Writing the observed hit rate as $\hat{p}=\hits/n$ turns \Cref{eq:ztest} into $z=\sqrt{n}\,(\hat{p}-\valfrac)/\sqrt{\valfrac(1-\valfrac)}$, so at a fixed hit rate above $\valfrac$ the score grows with $\sqrt{n}$.
Second, shorter units offer fewer interior points at which an attacker can split.
Third, a merge can become recoverable: because $\seg$ cuts on embedding change rather than punctuation, joining two marked \semspans into one sentence can preserve their original cut, allowing $\seg$ to re-cut near the provider's original point.

However, finer units require more provider draws from the LLM\@.
It samples candidate sentences as before, but commits only the first \semspan of the chosen candidate and regenerates the remainder at the next step, so a passage takes one best-of-$N$ round per \semspan\ rather than one per sentence.
Finer units repeat boundary-depth selection more often within each sentence, increasing its potential effect on marked-text fidelity.
\S\ref{subsec:eval-costs} measures the net fidelity change.

\textbf{Attacker response.}
Because $\seg$'s rule is public, the attacker instantiates it with $\Embsur$, applies it to marked text, and anchors on the recovered \semspans.
Finer units raise the attacker's cost as they raise the provider's: $\budget$ applies per unit, and a passage contains more \semspans than sentences.

\subsection{Scheme-Specific \eda}
\label{subsec:scheme-attack}

The three responses above compose into one attack, \edas, which runs \eda with the anchor rule and segmenter of its target rather than the fixed positional anchors of \edap.
We write each \edas\ configuration as (anchor rule, segmentation unit), with pos for positional anchoring and sent for sentence segmentation.
The public segmenter $\seg$ determines the source anchors, and the published valid-set rule determines how many anchors each candidate must move away from.

\begin{algorithm}[t]
\caption{Scheme-specific \eda\ (\edas).}
\label{alg:edas}
\begin{algorithmic}[1]
\Require marked text $x$; paraphraser $\pi$; surrogate encoder $\Embsur$; budget $\budget$; surrogate-bound segmenter $\seg$; valid-set rule
\Ensure attacked text $y=(u'_1,\dots,u'_{n'})$
\State $(a_1,\dots,a_n) \gets \seg(x)$ \label{line:manchors}
\For{$t = 1, 2, \dots$ \textbf{until} $\pi$ emits \eos\ or $y$ reaches \edamaxtok\ tokens}
  \State $\mathcal{C} = \{c_1,\dots,c_{\budget}\} \ni c_i \sim \pi(\,\cdot \mid x,\, u'_1,\dots,u'_{t-1},\, t)$ \label{line:msample}
  \If{valid-set rule is fixed} \label{line:mmode}
    \State $A_t \gets \{a_1,\dots,a_n\}$
  \Else
    \State $A_t \gets \{a_{\min(t,n)}\}$
  \EndIf
  \State $u'_t \gets \operatorname*{arg\,max}_{c \in \mathcal{C}} \ \min_{a \in A_t} d\!\big(\Embsur(c), \Embsur(a)\big)$ \label{line:mselect}
\EndFor
\State \Return $y = (u'_1,\dots,u'_{n'})$
\end{algorithmic}
\end{algorithm}

\Cref{alg:edas} instantiates \edas\ with the same paraphraser $\pi$, surrogate encoder $\Embsur$, and budget $\budget$ as \edap.
First, it applies the detector's $\seg$ to marked text and uses the recovered units as anchors $a_1,\dots,a_n$ (line~\ref{line:manchors}).
At output step $t$, it samples $\budget$ candidates from the paraphraser $\pi$ (line~\ref{line:msample}) and emits the candidate farthest from its nearest anchor in the active set $A_t$ (line~\ref{line:mselect}).
The watermark's valid-set rule selects $A_t$ (line~\ref{line:mmode}), recovering the fixed-set and context-dependent objectives defined above.
Either setting of $A_t$ combines with either segmenter, and best-of-$N$ changes neither attack input.
The attackers for the context-dependent \semstampfamily and \pmark\ use \edaconfigpossent, so \edas\ coincides with \edap.
The \samark\ attacker uses \edaconfigbagsent because its per-query key bit gives no role to source position (\S\ref{sec:background}).
The \ourstamp\ attacker and \kourstamp\ attacker use \edaconfigbagspan.

Every instantiation remains within the no-box threat model of \S\ref{sec:problem}: it uses the public segmenter and valid-set rule, but not a detector oracle or key material.
The surrogate encoder $\Embsur$ invokes the transfer assumption stated and tested in \S\ref{sec:attacks}.
Because the objective optimizes displacement rather than content preservation, a larger $\budget$ improves evasion only while $\pi$ still meets $\qbar$.
The appendix edit-signature analysis suggests that content-preserving \edas\ evasions combine reordering and resegmentation with rewording (\Cref{fig:attack-channels}).

%% file: figures/best_of_n.tex
\definecolor{bestNAccent}{HTML}{315D8A}

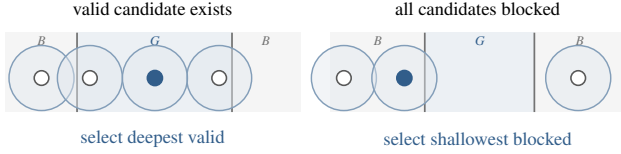
\begin{figure}[t]
  \centering
  \begin{tikzpicture}[
    x=1cm,
    y=1cm,
    candidate/.style={circle, draw=black!65, fill=white, line width=0.6pt,
                      minimum size=5.5pt, inner sep=0pt},
    selected/.style={candidate, draw=bestNAccent, fill=bestNAccent,
                     line width=0.8pt},
    attack/.style={draw=bestNAccent!60, fill=bestNAccent!14,
                   fill opacity=0.45, line width=0.55pt},
  ]
    \fill[black!4] (0,0.55) rectangle (0.95,1.60);
    \fill[bestNAccent!9] (0.95,0.55) rectangle (3.00,1.60);
    \fill[black!4] (3.00,0.55) rectangle (3.90,1.60);
    \draw[black!55, line width=0.65pt] (0.95,0.55) -- (0.95,1.60);
    \draw[black!55, line width=0.65pt] (3.00,0.55) -- (3.00,1.60);
    \node[font=\scriptsize] at (1.95,1.91) {valid candidate exists};
    \node[font=\tiny, text=black!55] at (0.48,1.49) {$B$};
    \node[font=\tiny, text=bestNAccent] at (1.98,1.49) {$G$};
    \node[font=\tiny, text=black!55] at (3.45,1.49) {$B$};

    \draw[attack] (0.48,1.00) circle[radius=0.43];
    \draw[attack] (1.12,1.00) circle[radius=0.43];
    \draw[attack] (2.83,1.00) circle[radius=0.43];
    \draw[attack] (1.98,1.00) circle[radius=0.43];
    \node[candidate] at (0.48,1.00) {};
    \node[candidate] at (1.12,1.00) {};
    \node[candidate] at (2.83,1.00) {};
    \node[selected] at (1.98,1.00) {};
    \node[font=\scriptsize, text=bestNAccent] at (1.95,0.20)
      {select deepest valid};

    \fill[black!4] (4.30,0.55) rectangle (5.55,1.60);
    \fill[bestNAccent!9] (5.55,0.55) rectangle (7.00,1.60);
    \fill[black!4] (7.00,0.55) rectangle (8.20,1.60);
    \draw[black!55, line width=0.65pt] (5.55,0.55) -- (5.55,1.60);
    \draw[black!55, line width=0.65pt] (7.00,0.55) -- (7.00,1.60);
    \node[font=\scriptsize] at (6.25,1.91) {all candidates blocked};
    \node[font=\tiny, text=black!55] at (4.93,1.49) {$B$};
    \node[font=\tiny, text=bestNAccent] at (6.28,1.49) {$G$};
    \node[font=\tiny, text=black!55] at (7.60,1.49) {$B$};

    \draw[attack] (4.48,1.00) circle[radius=0.43];
    \draw[attack] (5.28,1.00) circle[radius=0.43];
    \draw[attack] (7.58,1.00) circle[radius=0.43];
    \node[candidate] at (4.48,1.00) {};
    \node[selected] at (5.28,1.00) {};
    \node[candidate] at (7.58,1.00) {};
    \node[font=\scriptsize, text=bestNAccent] at (6.25,0.20)
      {select shallowest blocked};
  \end{tikzpicture}
  \caption{Disks illustrate displacement after a rewrite.
  Best-of-$N$ selects the deepest valid candidate if available (left), otherwise the shallowest blocked candidate (right).}
  \label{fig:best-of-n}
\end{figure}

%% file: figures/semspan.tex
\definecolor{spanAccent}{HTML}{315D8A}

\begin{figure}[t]
  \centering
  \begin{tikzpicture}[
    x=1cm, y=1cm,
    word/.style={draw=black!22, fill=black!3, rounded corners=1pt,
                 minimum width=0.80cm, minimum height=0.42cm,
                 inner sep=1pt, font=\scriptsize},
    tickok/.style={draw=black!55, line width=0.7pt},
    bar/.style={draw=none, fill=black!35},
    barmax/.style={draw=none, fill=spanAccent},
    span/.style={draw=spanAccent!70, fill=spanAccent!7, rounded corners=1.5pt,
                 line width=0.5pt},
    cutmark/.style={circle, draw=spanAccent, fill=white, line width=0.6pt,
                    inner sep=0.5pt, minimum size=7pt, font=\tiny\bfseries,
                    text=spanAccent},
  ]

    \draw[spanAccent!25, line width=0.5pt, dash pattern=on 1.5pt off 1.5pt]
      (3.28,-0.53) -- (3.28,2.12);

    \draw[black!20, line width=0.4pt] (0,1.50) -- (8.20,1.50);
    \fill[bar]    (2.38,1.50) rectangle (2.54,1.72);
    \fill[barmax] (3.20,1.50) rectangle (3.36,1.96);
    \fill[bar]    (4.84,1.50) rectangle (5.00,1.70);
    \fill[bar]    (5.66,1.50) rectangle (5.82,1.76);
    \node[font=\tiny, text=black!55]   at (2.46,1.83) {$.29$};
    \node[font=\tiny, text=spanAccent] at (3.28,2.07) {$.61$};
    \node[font=\tiny, text=black!55]   at (4.92,1.81) {$.26$};
    \node[font=\tiny, text=black!55]   at (5.74,1.87) {$.34$};

    \foreach \i/\txt in {1/The, 2/reactor, 3/heats, 4/water, 5/and,
                         6/rotors, 7/convert, 8/steam, 9/to, 10/power}
      \node[word] at ({(\i-0.5)*0.82}, 0.85) {\txt};

    \foreach \i in {3,6,7}
      \draw[tickok] ({\i*0.82},0.56) -- ({\i*0.82},1.14);
    \draw[spanAccent, line width=1.1pt] (3.28,0.52) -- (3.28,1.18);

    \draw[decorate, decoration={brace, amplitude=3pt, mirror, raise=1pt},
          black!45, line width=0.4pt] (0.82,0.55) -- (3.24,0.55)
      node[midway, below=7pt, font=\tiny, text=black!55] {$\spanwin$ words};
    \draw[decorate, decoration={brace, amplitude=3pt, mirror, raise=1pt},
          black!45, line width=0.4pt] (3.32,0.55) -- (5.74,0.55)
      node[midway, below=7pt, font=\tiny, text=black!55] {$\spanwin$ words};

    \node[cutmark] at (3.28,1.30) {1};
    \node[cutmark] at (5.74,1.30) {2};
    \draw[span] (0.02,-0.53) rectangle (3.26,-0.20);
    \draw[span] (3.30,-0.53) rectangle (5.72,-0.20);
    \draw[span] (5.76,-0.53) rectangle (8.18,-0.20);
    \node[font=\scriptsize, text=spanAccent] at (1.64,-0.365) {$u_1$};
    \node[font=\scriptsize, text=spanAccent] at (4.51,-0.365) {$u_2$};
    \node[font=\scriptsize, text=spanAccent] at (6.97,-0.365) {$u_3$};
    \node[anchor=north, font=\tiny, text=black!55] at (1.64,-0.59)
      {kept: $4\le\spanmax$};
    \node[anchor=north, font=\tiny, text=black!55] at (5.74,-0.59)
      {re-cut: $6>\spanmax$};

  \end{tikzpicture}
  \caption{$\seg$ segmentation ($\spanwin=3$, $\spanmax=5$).
  Ticks mark valid cuts; bars give window distances.
  $\seg$ cuts at the largest distance~(1), then recursively cuts the six-word right \semspan~(2).}
  \label{fig:semspan}
\end{figure}
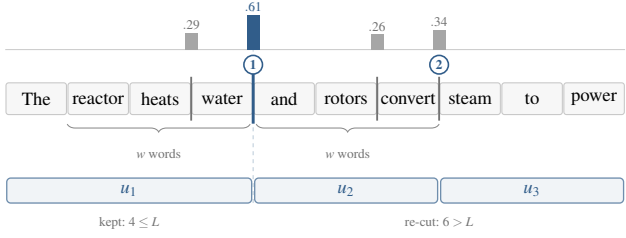

%% file: sections/evaluation.tex
\section{Robustness, Fidelity, and Cost}
\label{sec:eval}
Our evaluation determines whether each design component resists its target edit type and measures the provider cost of that resistance.
We attribute score loss to individual edit types, test removal under no-box and detector-access conditions, and quantify clean detection and generation costs.
\input{tables/costs}
\input{figures/channel_response}

\subsection{Experimental Setup}
\label{subsec:eval-setup}

\textbf{Watermarks.}
We use each watermark's published configuration.
\semstamp\ and \ksemstamp\ use $\valfrac=0.25$, $\nregions$ regions, and margins of $0.02$ and $0.035$, respectively.
We evaluate online \pmark\ with $b=4$ channels and \samark\ with $b=2$ channels.
Every scheme may draw at most $\Nprov=\nprov$ candidates per unit.
Rejection sampling may stop at its first accepted candidate, while best-of-$N$, \samark, and \pmark\ use the full pool.
For \semspan\ segmentation, we set the maximum unit length to $\spanmax=15$ words and the boundary-comparison window to $\spanwin=5$ words on each side.
For each $k$-means configuration, we recompute the centroids from its sentence or \semspan embeddings over a held-out corpus of $8\,192$ C4 documents~\cite{raffel2020c4}, following \ksemstamp's fitting procedure~\cite{hou2024ksemstamp}.

\textbf{Models and data.}
The provider uses \llama~\cite{grattafiori2024llama3}.
The provider-side \semstampfamily and \ourstampfamily configurations use the released C4-tuned \texttt{AbeHou/SemStamp-c4-sbert} checkpoint, while \pmark\ and \samark\ use \texttt{all-mpnet-base-v2}~\cite{reimers2019sentencebert}.
We generate every configuration from one set of $\ndocs$ prompts from C4's \texttt{realnewslike} configuration~\cite{raffel2020c4}, stopping at $\maxnewtok$ new tokens or, for \pmark\ and \samark, targeting $12$ sentences.
A disjoint $\nnull$-document split supplies the empirical null for every configuration except online \pmark, after truncation under the scheme's stopping rule to match the marked-document length distribution.
We set the $5\%$ and $1\%$ thresholds to the empirical $0.95$ and $0.99$ score quantiles, respectively; online \pmark instead uses the nominal standard-normal thresholds $1.6449$ and $2.3263$.

\textbf{Attacks.}
Our sentence-level controls are \pegasus~\cite{zhang2020pegasus} and \parrot~\cite{damodaran2021parrot}, each with and without bigram selection.
Bigram selection chooses the lowest-overlap candidate whose BERTScore~\cite{zhang2020bertscore} is within $3\%$ of the best candidate.
For \dipper~\cite{krishna2023paraphrasing}, we fix lexical diversity at $20$ while sweeping order diversity over $\{20,40,80\}$, then fix order diversity at $20$ while sweeping lexical diversity over $\{20,40,80\}$.
For no-box \eda, we use \texttt{Qwen2.5-3B-Instruct}~\cite{yang2024qwen25} and a \texttt{bge-base-en-v1.5} surrogate encoder~\cite{xiao2023cpack} rather than the provider encoder.
This gives the attacker a smaller generator and a surrogate with a different architecture from every provider encoder.
This separation tests whether \eda's displacement objective transfers without reusing the provider's generator or encoder.
The \edas\ budget grid is $\{1,2,4,8,16,32,64\}$.
Additional \edap\ comparisons use available budgets in $\{4,16,64\}$, and \edad\ adds $\{8,32\}$.

\textbf{Detection and quality.}
We use $\ASR$@\resultPrimaryFpr\ as the primary operating point and sweep $\qbar$ from $\qbarmin$ through $\qbarmax$ in steps of $\qbarstep$.
We report the clean true-positive rate ($\TPR$), the fraction of unattacked marked documents flagged at the same threshold.
At each $\qbar$, we select the setting from each attack's effort grid that maximizes $\ASR$.
\qwen~\cite{yang2025qwen3} judges fidelity from fluency, coherence, relevance, and informativeness, and content preservation from content recall, detail precision, information injection, and contradiction.
\Cref{app:eval-judge} specifies both judges, their prompts, and the scoring procedure.
We run each judgment with three random seeds and average the scores.
Both scores are proxies for human judgments.

\textbf{Edit measurement.}
For a controlled edit $\mathcal{A}$ applied to marked documents $x_1,\ldots,x_D$, normalized $z$ retention is
\begin{equation}
  \zret(\mathcal{A})
  = \frac{\sum_{i=1}^{D} z_k(\mathcal{A}(x_i))}{\sum_{i=1}^{D} z_k(x_i)},
  \label{eq:z-retention}
\end{equation}
where $D$ is the number of documents and $k$ is the detector key.
We compute paired-bootstrap intervals over documents.
For marked text $x$ and attacked text $y$, let $w_1,\ldots,w_m$ be patience anchors, tokens that occur exactly once in each text, ordered by their positions in $x$~\cite{heckel1978diff}.
Anchor $w_j$ has source and attacked token positions $(p_j,p'_j)$ and sentence indices $(u_j,u'_j)$ under NLTK's pretrained English Punkt sentence tokenizer~\cite{loper2002nltk}.
Let $\mathcal{P}=\{(i,j):i<j, u_i\neq u_j\}$ be the cross-sentence anchor pairs.
We define $e_{\mathrm{merge}}$ as the fraction of adjacent source-order anchors separated in $x$ but joined in $y$, and $e_{\mathrm{split}}$ as the fraction of adjacent attacked-order anchors joined in $x$ but separated in $y$.
We define the resulting edit rates as follows.
\begin{equation}
  \begin{aligned}
    r_{\mathrm{word}} &= 1-F_1^{\mathrm{ROUGE\text{-}1}}(x,y), \\
    r_{\mathrm{order}} &= \frac{1}{|\mathcal{P}|}\sum_{(i,j)\in\mathcal{P}}
      \mathbf{1}\!\left[(p_i-p_j)(p'_i-p'_j)<0\right], \\
    r_{\mathrm{seg}} &= 1-
      \frac{2(1-e_{\mathrm{merge}})(1-e_{\mathrm{split}})}
      {(1-e_{\mathrm{merge}})+(1-e_{\mathrm{split}})}.
  \end{aligned}
  \label{eq:channel-measures}
\end{equation}
The rates are $1-$ROUGE-1 F1 for rewording~\cite{lin2004rouge}, a restricted Kendall inversion rate for reordering~\cite{kendall1938rank}, and $1-$ boundary F1 for resegmentation~\cite{beeferman1999segmentation,pevzner2002segmentation}.

\textbf{Component sequence.}
\Cref{tab:costs} lists the five-stage LSH and $k$-means sequences from \semstampfamily to \ourstampfamily and reports each stage's \edas\ configuration as (anchor rule, segmentation unit).
Because each attacker uses the public anchor rule and segmenter at that stage, each measured change accounts for an adversary that knows the added component.
The diversity ranking targets no edit type, so we measure its effects on retention, fidelity, and repetition rather than evaluate a designated defense.

\subsection{Score Loss by Edit Type}
\label{subsec:eval-channels}
\Cref{fig:channel-response} reports normalized score retention, $\zret$, against the measured edit rates of in-place synonym substitution, sentence reordering, and resegmentation.
\Cref{tab:isolation} reports that each edit changes its target rate while leaving the other two near zero.
\ksemstamp loses retention under all three controlled edits, including structural edits that preserve sentence text or claims.
This loss supports the mechanism in \S\ref{sec:attacks}: the detector rebuilds state from attacker-controlled structure.

Best-of-$N$ raises retention under rewording, but it does not eliminate rewording-induced score loss.
The fixed valid set removes predecessor re-keying and eliminates reordering-induced score loss.
The diversity ranking lowers rewording and resegmentation retention relative to the fixed-set stage, but improves fidelity and reduces repetition in both component sequences (\Cref{tab:costs}).
Adding \semspans then raises resegmentation retention above every sentence-based stage.

\subsection{No-Box Robustness}
\label{subsec:eval-robustness}
\input{figures/cross_scheme}

\Cref{fig:cross-scheme} reports the strongest tested no-box attack family for each scheme and the effect of configuring \eda with the scheme's public anchor rule and segmenter.
Each panel traces the strongest setting for sentence-level controls, \dipper, and \edas\ across $\qbar$ under the no-box threat model.
For \samark, \ourstamp, and \kourstamp, dashed \edap\ curves compare positional sentence anchors with bag anchoring.

At $\qbar=\costasrqualitybar$, \edas\ $\ASR$ falls from \resultSemEdasAsr\ to \resultSwordEdasAsr\ on LSH and from \resultKSemEdasAsr\ to \resultKSwordEdasAsr\ on $k$-means.
At the same $\qbar$, \edas's $\ASR$ estimate is $3.0$ percentage points higher than \dipper's on \ksemstamp\ and is higher on every other scheme in \Cref{fig:cross-scheme}.
The \ourstamp\ and \kourstamp\ attackers replace \edap's positional sentence anchors with bag \semspan\ anchors.
Using bag \semspan\ anchors raises $\ASR$ from \resultSwordEdapAsr\ to \resultSwordEdasAsr\ on \ourstamp\ and from \resultKSwordEdapAsr\ to \resultKSwordEdasAsr\ on \kourstamp.
The \samark\ attacker replaces only positional anchors with bag anchors and continues to use sentence units.
Its \resultSamarkEdasAsr\ $\ASR$ effectively coincides with \edap's \resultSamarkEdapAsr\ at this $\qbar$, although the \edas attacker leads at every other $\qbar$.

\subsection{Robustness under Detector Access}
\label{subsec:eval-informed}

\emph{Detector-access EDA} (\edad) augments \edap\ with the provider's exact candidate-level region test, including its key, encoder, partition, and segmenter.
It retains \edap's paraphraser and per-unit search budget, and it does not receive the provider's generator or quality judge.
We evaluate this attacker on the $k$-means pair, \ksemstamp\ and \kourstamp.
For these schemes, let $h_k(c\mid p,t)\in\{0,1\}$ denote the detector's decision for candidate $c$ at position $t$ under committed predecessor $p$, with $1$ for a hit and $0$ for a miss.
At each output step, \edad\ selects the detector miss closest to its source anchor when one exists; otherwise, it selects the most displaced candidate as in \edap.
\Cref{alg:detector-access} specifies the procedure.
\Cref{fig:whitebox} reports raw evasion, $\ASR$@\resultPrimaryFpr\ at $\qbar=\costasrqualitybar$, and the proportion of detector-evading outputs that meet $\qbar$ as the detector-access search budget increases for the $k$-means schemes.

\begin{algorithm}[t]
\caption{Detector-access \eda\ (\edad).}
\label{alg:detector-access}
\begin{algorithmic}[1]
\Require marked text $x$; paraphraser $\pi$; provider detector $\Detect(k,\cdot)$ with encoder $\Emb$, partition $P$, and segmenter $\seg$; budget $\budget$
\Ensure attacked text $y=(u'_1,\dots,u'_{n'})$
\State $(a_1,\dots,a_n) \gets \seg(x)$; $u'_0 \gets \varnothing$ \label{line:dseg}
\For{$t = 1, 2, \dots$ \textbf{until} $\pi$ emits \eos\ or $y$ reaches \edamaxtok\ tokens}
  \State $\mathcal{C} = \{c_1,\dots,c_{\budget}\} \ni c_i \sim \pi(\,\cdot \mid x,\, u'_1,\dots,u'_{t-1},\, t)$ \label{line:dsample}
  \State $a \gets a_{\min(t,n)}$
  \State $\mathcal{C}_{\mathrm{miss}} \gets \{c \in \mathcal{C}: h_k(c \mid u'_{t-1},t)=0\}$ \label{line:dmiss}
  \If{$\mathcal{C}_{\mathrm{miss}} \neq \varnothing$}
    \State $u'_t \gets \operatorname*{arg\,min}_{c \in \mathcal{C}_{\mathrm{miss}}} d\!\big(\Emb(c),\Emb(a)\big)$ \label{line:dselect}
  \Else
    \State $u'_t \gets \operatorname*{arg\,max}_{c \in \mathcal{C}} d\!\big(\Emb(c),\Emb(a)\big)$ \label{line:dfallback}
  \EndIf
\EndFor
\State \Return $y = (u'_1,\dots,u'_{n'})$
\end{algorithmic}
\end{algorithm}

At $\budget=4$, raw evasion is \resultWhiteboxKSemEvasionLowBudget\ on \ksemstamp\ and \resultWhiteboxKSwordEvasionLowBudget\ on \kourstamp.
More search raises raw evasion against both schemes, but the selected rewrites against \kourstamp\ receive progressively lower content-preservation scores.
After quality gating, $\ASR$ is \resultWhiteboxKSwordAsrRange\ on \kourstamp, compared with \resultWhiteboxKSemAsrRange\ on \ksemstamp.
\input{figures/whitebox}

\subsection{Clean Performance and Costs}
\label{subsec:eval-costs}
At \resultPrimaryFpr\ FPR, clean $\TPR$ rises from \resultSemCleanTpr\ to \resultSwordCleanTpr\ on LSH and remains effectively unchanged from \resultKSemCleanTpr\ to \resultKSwordCleanTpr\ on $k$-means (\Cref{tab:costs}).
Fidelity falls from \resultSemFidelity\ to \resultSwordFidelity\ on LSH and from \resultKSemFidelity\ to \resultKSwordFidelity\ on $k$-means.

\pmark\ has the highest measured fidelity and is the strongest scheme against the sentence-level controls, consistent with its reported performance~\cite{huo2025pmark}.
Against the whole-text \dipper\ and \edas, both \ourstampfamily configurations yield lower $\ASR$ than \pmark.
\kourstamp\ yields lower attack success than \ourstamp, while \ourstamp\ retains higher fidelity.

The component sequence attributes the robustness and fidelity trade-offs.
Best-of-$N$ raises clean $\TPR$ under both partition methods and reduces $\ASR$ for every no-box attack, at a larger fidelity cost on $k$-means.
The fixed valid set reduces $\ASR$ against \dipper\ and \edas, but it requires higher detection thresholds under both partition methods.

Among the four comparison schemes, \samark\ has the lowest \edas\ $\ASR$, \resultSamarkEdasAsr\ at \resultPrimaryFpr\ FPR, but also the lowest clean $\TPR$ at both reported operating points.
Its self-anchored score infers each key bit from the suspect text (\S\ref{sec:background}), so a topically concentrated passage can receive a high score whether or not it carries a mark.
At \resultPrimaryFpr\ FPR, calibration sets the threshold to \resultSamarkThreshold, and \samark\ flags \resultSamarkCleanTpr\ of unattacked marked documents.
At $1\%$ FPR, its clean $\TPR$ falls to \resultSamarkCleanTprOne.
For comparison, \ourstamp\ pairs \resultSwordCleanTpr\ clean $\TPR$ with \resultSwordEdasAsr\ \edas\ $\ASR$, and \kourstamp\ pairs \resultKSwordCleanTpr\ with \resultKSwordEdasAsr.

\Semspans lower \edas\ success under both partition methods.
Overall, the fixed valid set and \semspans supply most robustness; the diversity ranking trades some robustness for fidelity and lower repetition.

%% file: tables/costs.tex
\begin{table*}[t]
  \centering
  \scriptsize
  \setlength{\tabcolsep}{3pt}
  \caption{%
    Clean detection, generation cost, output quality, and robustness at $\Nprov=\nprov$.
    \edas\ configurations list (anchor rule, unit): pos and bag denote anchor rules, and sent and span denote sentence and \semspan\ units.
    Swords mark final designs.
    At $\qbar=\costasrqualitybar$, $\ASR$ cells report the strongest judged setting per attack family at $1\%$ and $5\%$ FPR\@.
    PPL, Sem.\ entropy, Rep-3, and Sent.\ dup.\ denote corpus perplexity, semantic entropy, repeated-trigram rate, and sentence duplication.
    Rates are percentages; fidelity reports mean [$95\%$ CI], and Draws/sentence reports mean candidate draws per clean sentence.
    Bold and underlined values mark the overall best and worst means.%
  }
  \label{tab:costs}
  \begin{tabular}{@{}lcccccc@{\hspace{4pt}}cccccc@{}}
    \toprule
    \multirow[b]{2}{*}{Configuration} &
      \multirow[b]{2}{*}{\shortstack[c]{\edas\\config.}} &
      \multicolumn{2}{c}{@$1\%$ / $5\%$} &
      \multicolumn{3}{c}{$\ASR\,\downarrow$ (\%) by no-box attacker, @$1\%$ / $5\%$} &
      \multirow[b]{2}{*}{\shortstack[c]{Fidelity $\uparrow$\\{[95\% CI]}}} &
      \multirow[b]{2}{*}{\shortstack[c]{Draws/\\sent. $\downarrow$}} &
      \multirow[b]{2}{*}{PPL $\downarrow$} &
      \multirow[b]{2}{*}{\shortstack[c]{Sem.\\entropy $\uparrow$}} &
      \multirow[b]{2}{*}{Rep-3 $\downarrow$} &
      \multirow[b]{2}{*}{\shortstack[c]{Sent.\\dup.\ $\downarrow$}} \\
    \cmidrule(lr){3-4}\cmidrule(lr){5-7}
    & & $\TPR\,\uparrow$ & $\thresh$ &
      Sentence & \dipper & \edas &
      & & & & & \\
    \midrule
    \publishedcostrows
    \midrule
    \lshladdercostrows
    \midrule
    \kmeansladdercostrows
    \bottomrule
  \end{tabular}
\end{table*}

%% file: figures/channel_response.tex
\begin{figure*}[t]
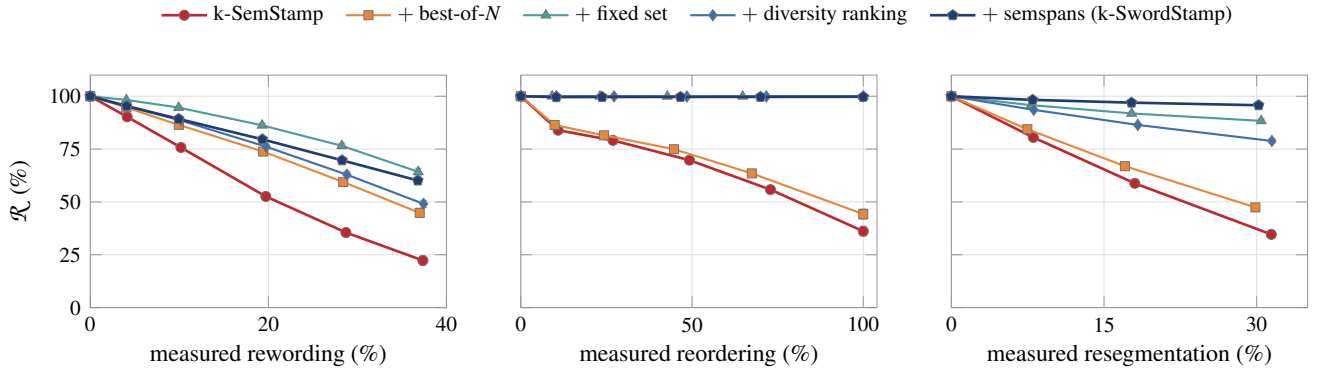

  \centering
  \chanrespbody{\rungprefix}{\ladderbase}{\ladderfinal}
  \caption{%
    Normalized detection-score retention $\zret$ versus realized edit strength, with bootstrap $95\%$ intervals.
    Resegmentation pools sentence splits and merges using boundary disagreement.
    Lines follow the five stages from \ladderbase\ to \ladderfinal.%
  }
  \label{fig:channel-response}
\end{figure*}

%% file: figures/cross_scheme.tex
\newcommand{\frontierplot}[2]{%
  \IfFileExists{plots/data/cross_scheme/#1-#2.dat}{%
    \dataplot [frontier/#2, error bars/.cd, y dir=both, y explicit]
      table [x=bar, y=y, y error minus=em, y error plus=ep]
      {plots/data/cross_scheme/#1-#2.dat};%
  }{}%
}

\newcommand{\frontierpartialplot}[2]{%
  \IfFileExists{plots/data/cross_scheme/#1-#2.dat}{%
    \dataplot [frontier/#2, frontier/partial, error bars/.cd,
               y dir=both, y explicit]
      table [x=bar, y=y, y error minus=em, y error plus=ep]
      {plots/data/cross_scheme/#1-#2.dat};%
  }{}%
}

\newcommand{\frontierlabels}[2]{%
  \IfFileExists{plots/data/cross_scheme/#1-#2-labels.dat}{%
    \dataplot [frontier/label/#2, nodes near coords,
              point meta=explicit symbolic]
      table [x=bar, y=y, meta=label]
      {plots/data/cross_scheme/#1-#2-labels.dat};%
  }{}%
}

\newcommand{\frontierallabels}[1]{%
  \frontierlabels{#1}{sentence}%
  \frontierlabels{#1}{dipper}%
  \frontierlabels{#1}{adaptive}%
}

\newcommand{\frontiernobox}[2]{%
  #2{#1}{sentence}%
  #2{#1}{dipper}%
  #2{#1}{adaptive-base}%
  #2{#1}{adaptive}%
}

\newcommand{\frontierall}[1]{%
  \frontiernobox{#1}{\frontierplot}%
  \frontierallabels{#1}%
}

\newcommand{\frontierpartial}[1]{%
  \frontiernobox{#1}{\frontierpartialplot}%
  \frontierallabels{#1}%
  \frontierpending{#1}%
}

\newcommand{\frontierpanel}[1]{%
  \IfFileExists{plots/data/cross_scheme/#1-adaptive.dat}{%
    \IfFileExists{plots/data/cross_scheme/#1-dipper.dat}{%
      \frontierall{#1}%
    }{\frontierpartial{#1}}%
  }{\frontierpartial{#1}}%
}

\newcommand{\frontierpending}[1]{%
  \IfFileExists{plots/data/cross_scheme/#1-adaptive.dat}{}{%
    \node[align=center, font=\scriptsize, text=black!55]
      at (rel axis cs:0.50,0.50) {adaptive\\pending};%
  }%
  \IfFileExists{plots/data/cross_scheme/#1-dipper.dat}{}{%
    \node[align=center, font=\scriptsize, text=black!55]
      at (rel axis cs:0.50,0.34) {Dipper\\pending};%
  }%
}

\pgfplotsset{
  frontier base/.style={
    scale only axis,
    qbar axis,
    asr axis,
    ymax=0.85,
    ytick={0,0.2,0.4,0.6,0.8},
    width=0.27\textwidth, height=\plotpanelheight,
    title style={font=\small, yshift=-2pt},
  },
  frontier group/.style={
    group/group size=3 by 2,
    group/horizontal sep=23pt,
    group/vertical sep=17pt,
    group/xlabels at=edge bottom,
    group/xticklabels at=edge bottom,
    group/ylabels at=edge left,
    group/yticklabels at=edge left,
  },
  frontier legend/.style={
    legend style={
      at={(0.5,1.24)}, anchor=south,
      legend columns=4, draw=none, font=\footnotesize,
      cells={anchor=west},
      /tikz/every even column/.append style={column sep=5pt},
    },
  },
}

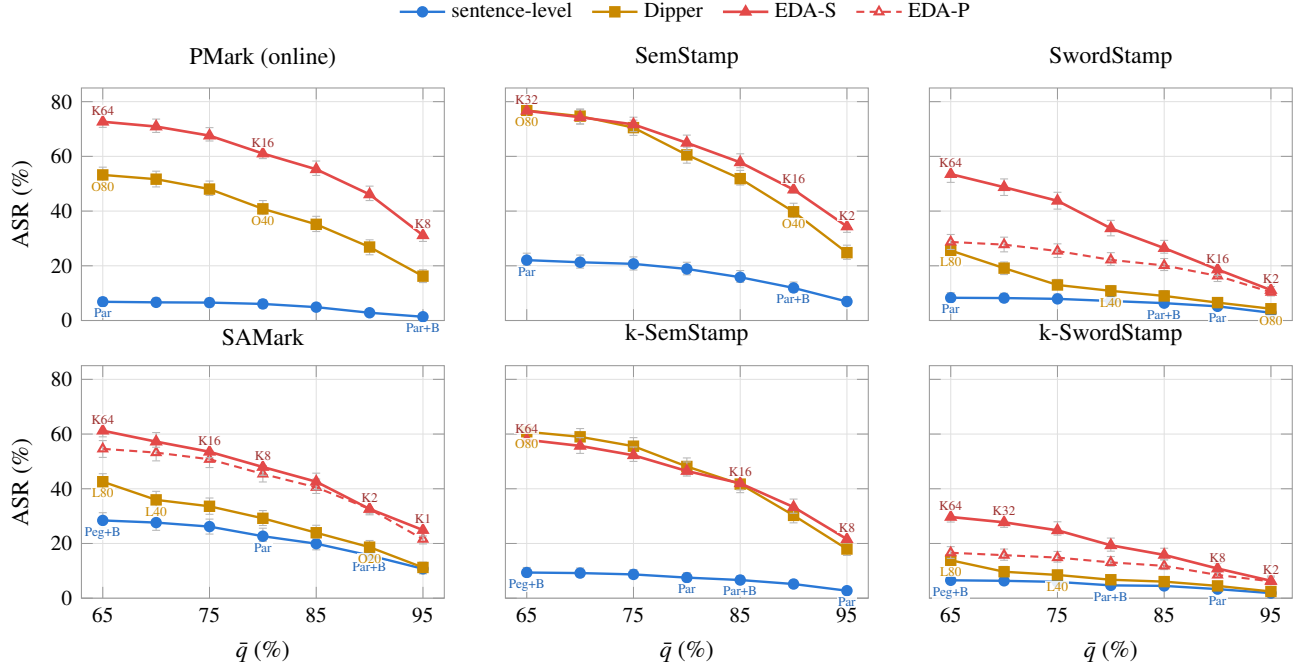
\begin{figure*}[t]
  \centering
  \begin{tikzpicture}
    \begin{groupplot}[frontier base, frontier group]
      \nextgroupplot[title={\pmark\ (online)}]
      \frontierpanel{pmark-online}

      \nextgroupplot[title={\semstamp}, frontier legend]
      \addlegendimage{frontier/sentence}
      \addlegendentry{sentence-level}
      \addlegendimage{frontier/dipper}
      \addlegendentry{\dipper}
      \addlegendimage{frontier/adaptive}
      \addlegendentry{\edas}
      \addlegendimage{frontier/adaptive-base}
      \addlegendentry{\edap}
      \frontierpanel{semstamp}

      \nextgroupplot[title={\ourstamp}]
      \frontierpanel{firmstamp}

      \nextgroupplot[title={\samark}]
      \frontierpanel{samark}

      \nextgroupplot[title={\ksemstamp}]
      \frontierpanel{ksemstamp}

      \nextgroupplot[title={\kourstamp}]
      \frontierpanel{kfirmstamp}
    \end{groupplot}
  \end{tikzpicture}
  \caption{%
    No-box ASR at \resultPrimaryFpr\ FPR versus $\qbar$, with paired-document bootstrap $95\%$ intervals.
    Each curve selects the strongest setting per attack family and $\qbar$.
    Labels identify sentence-level controls (Peg, Par, $+$B), \dipper\ settings (O, L), and \eda\ budgets (K).
    Dashed curves show \edap\ at $\budget\in\{4,16,64\}$.%
  }
  \label{fig:cross-scheme}
\end{figure*}

%% file: figures/whitebox.tex
\newcommand{\whiteboxseries}[3]{%
  \dataplot [#1, error bars/.cd, y dir=both, y explicit]
    table [x=K, y=y, y error minus=em, y error plus=ep]
    {plots/data/whitebox/#2-#3.dat};%
}

\pgfplotsset{
  whitebox base/.style={
    paper axis,
    width=0.39\textwidth, height=\plotpanelheight,
    scale only axis,
    xmode=log, log basis x=2,
    xmin=3.5, xmax=70,
    xtick={4,8,16,32,64},
    xticklabels={$4$,$8$,$16$,$32$,$64$},
    xlabel={per-unit search budget $\budget$},
    title style={font=\small, yshift=-3pt},
    error bars/error bar style={line width=0.4pt, draw=black!28},
    error bars/error mark options={rotate=90, mark size=1.5pt,
                                   draw=black!28, line width=0.4pt},
    clip mode=individual,
  },
  whitebox outcomes/.style={
    fraction y as percent,
    ymin=0.15, ymax=1.0,
    ytick={0.2,0.4,0.6,0.8,1.0},
    ylabel={outcome rate (\%)},
  },
  whitebox quality pass/.style={
    fraction y as percent,
    ymin=0.4, ymax=0.8,
    ytick={0.4,0.5,0.6,0.7,0.8},
    ylabel={quality-pass rate (\%)},
  },
  whitebox/asr/.style={densely dashed},
  whitebox legend/.style={
    legend style={
      at={(1.12,1.23)}, anchor=south,
      legend columns=2, draw=none, font=\footnotesize,
      cells={anchor=west},
      /tikz/every even column/.append style={column sep=7pt},
    },
  },
}

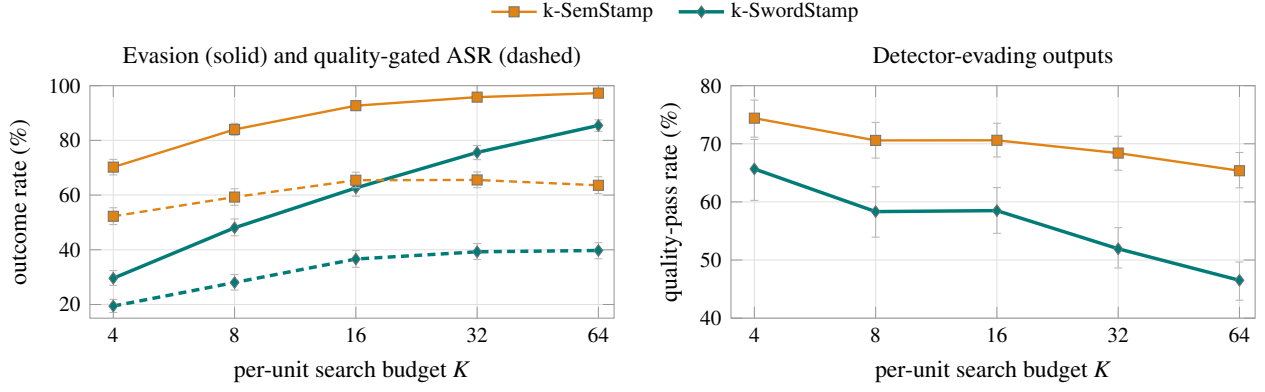
\begin{figure*}[t]
  \centering
  \begin{tikzpicture}
    \begin{groupplot}[
      whitebox base,
      group style={
        group size=2 by 1,
        horizontal sep=44pt,
      },
    ]
      \nextgroupplot[title={Evasion (solid) and quality-gated $\ASR$ (dashed)}, whitebox outcomes, whitebox legend]
      \addlegendimage{scheme/ksemstamp}
      \addlegendentry{\ksemstamp}
      \addlegendimage{scheme/kfirmstamp}
      \addlegendentry{\kourstamp}
      \whiteboxseries{scheme/ksemstamp}{evasion}{ksemstamp}
      \whiteboxseries{scheme/kfirmstamp}{evasion}{kfirmstamp}
      \whiteboxseries{scheme/ksemstamp, whitebox/asr}{asr}{ksemstamp}
      \whiteboxseries{scheme/kfirmstamp, whitebox/asr}{asr}{kfirmstamp}

      \nextgroupplot[title={Detector-evading outputs}, whitebox quality pass]
      \whiteboxseries{scheme/ksemstamp}{quality-pass-given-evasion}{ksemstamp}
      \whiteboxseries{scheme/kfirmstamp}{quality-pass-given-evasion}{kfirmstamp}
    \end{groupplot}
  \end{tikzpicture}
  \caption{%
    Detector-access \eda\ (\edad) against \ksemstamp\ and \kourstamp\ at \resultPrimaryFpr\ FPR\@.
    Left: raw evasion (solid) and $\ASR$ at $\qbar=\costasrqualitybar$ (dashed); right: fraction of evasions meeting the same $\qbar$.
    Bars give paired-document bootstrap $95\%$ intervals.%
  }
  \label{fig:whitebox}
\end{figure*}

%% file: sections/discussion.tex
\section{Limitations and Future Work}
\label{sec:discussion}
\textbf{Generation cost.}
Best-of-$N$ evaluates all $\Nprov=\nprov$ candidates, whereas rejection sampling stops after the first accepted candidate and uses fewer draws on average in our evaluation.
The fixed pool can be generated and scored in one batch, while rejection sampling makes acceptance decisions sequentially.
\Semspan generation adds sequential rounds because the provider commits one \semspan at a time.
The final LSH and $k$-means designs respectively average $1.6$ and $1.7$ $\nprov$-candidate rounds per output sentence (\Cref{tab:costs}).

\textbf{Evaluation scope.}
Our evaluation covers English news continuations up to \maxnewtok\ tokens from one provider model.
It does not establish whether the same attacks and defenses transfer to code generation, question answering, instruction following, other languages, longer documents, or other model families.
These settings may require different segmentation rules and definitions of content preservation.
Within our setting, we test two provider encoders but only one \eda paraphraser and one surrogate encoder.
We also rely on model-based fidelity and content-preservation scores rather than human judgments.

\textbf{Future work.}
Neither \eda\ nor \edad\ uses a quality oracle during candidate selection.
A quality-aware attacker could reject bad rewrites and achieve a different ASR-content-preservation tradeoff.
A provider could also use task-specific quality scores during candidate selection to recover part of the measured fidelity loss.
On the provider side, since the valid set is fixed across positions, the watermark is a static signal and can be learned directly into the generator.
The provider could therefore fine-tune its generator to generate watermarked samples from a smaller candidate pool.

%% file: sections/related_work.tex
\section{Related Work}
\label{sec:related}

\textbf{Watermarking versus post-hoc detection.}
Post-hoc detection is the main alternative to proactive watermarking.
Its detectors come as zero-shot statistical tests, such as the probability-curvature test of Mitchell et al.~\cite{mitchell2023detectgpt}, and as trained classifiers.
Neither requires generator cooperation, but both rely on empirical error estimates and rarely reach the low FPRs needed for deployment~\cite{zhao2025sok}.
Watermarking instead uses a keyed signal during generation to control FPR~\cite{zhao2025sok}.

\textbf{Watermark families.}
Token-level watermarks embed the signal in the token distribution and therefore tie it to specific token choices, which paraphrasing can erase~\cite{kirchenbauer2023watermark,zhao2024provable,aaronson2023watermarking,kuditipudi2023robust,christ2024undetectable,dathathri2024synthid}; Zhao et al.~\cite{zhao2025sok} survey that family in full.
Beyond the \ebws we evaluate, other schemes vary how the per-sentence signal is defined and selected, through similarity intervals~\cite{dabiriaghdam2025simmark}, embedding rings~\cite{ren2023semamark}, structure and cohesion partitions~\cite{zhang2024personamark,zhang2025cohemark}, and sparse-autoencoder features~\cite{yu2025saemark}.
\samark~\cite{huo2026samark} is concurrent work that uses per-query key bits and self-anchoring to make detection order-robust; we report its detection rate, calibrated threshold, and robustness alongside every other configuration in \S\ref{sec:eval}.

\textbf{Candidate selection and distortion.}
Selecting among sampled candidates to strengthen a mark predates our use of it.
WaterMax~\cite{giboulot2024watermax} and the black-box scheme of Bahri et al.~\cite{bahri2024blackbox} score chunks of generated text and keep the most detectable one, trading generation cost for detection strength at the token level.
Our best-of-$\Nprov$ selection makes the same trade at the sentence or \semspan level, scoring each candidate by how far inside its valid region it falls (\S\ref{subsec:best-of-n}).
Such selection distorts the generator's output distribution; distortion-free schemes avoid this effect by construction~\cite{kuditipudi2023robust,christ2024undetectable,hu2023unbiased}, and \pmark\ carries that property into the \ebw family~\cite{huo2025pmark}.

\textbf{Attacks and robustness evaluations.}
Prior evaluations of text watermarks center on paraphrasing.
Krishna et al.~\cite{krishna2023paraphrasing} show that a trained whole-text paraphraser evades token-level detectors unless the provider retains every generation for retrieval-based detection.
Hou et al.~\cite{hou2024semstamp,hou2024ksemstamp} report that \semstampfamily schemes resist sentence-by-sentence paraphrasers, and name whole-text paraphrase as the boundary of that result.
We test whole-text paraphrasing while holding the attacker's capabilities, $\qbar$, and each detector's operating point fixed across the family, and we scale attacker effort instead of fixing one paraphraser (\S\ref{sec:eval}).
Diaa et al.~\cite{diaa2024textwatermark} tune paraphrasers against a known token-level watermark, which is the closest prior attacker to ours in capability.
Other attacks assume access we deny: watermark stealing and design probing collect many keyed samples~\cite{jovanovic2024watermark,pang2024no}, while \textsc{De-mark} queries the watermarked generator to recover and remove n-gram watermarks~\cite{chen2025demark}.
Our attacker sees only the target text (\S\ref{sec:problem}).

\textbf{Impossibility results.}
Prior impossibility results address post-hoc detection and watermarking under different assumptions.
Sadasivan et al.~\cite{sadasivan2023aigenerated} and Chakraborty et al.~\cite{chakraborty2023possibilities} debate whether post-hoc detection survives as model and human text converge.
Zhang et al.~\cite{zhang2024watermarks} study an attacker with a quality oracle that checks candidate outputs and a perturbation oracle that proposes edits with a nontrivial chance of preserving quality.
They prove that the resulting random walk removes the watermark when it mixes rapidly over high-quality outputs.
Our no-box attacker has neither oracle, uses a generator weaker than the provider's, and has a finite search budget, so our results cover only the attackers and budgets we test, not Zhang et al.'s stronger setting.

%% file: sections/conclusion.tex
\section{Conclusion}
\label{sec:conclusion}
We measure how rewording, reordering, and resegmentation affect encoder-based watermark detection.
We consolidate them into \eda, a single adaptive no-box attack guided by embedding displacement.
At \resultPrimaryFpr\ FPR, \eda removes the watermark while retaining a \resultQualityRequirement\ content-preservation score for \resultPriorEdapAsrRange\ of outputs from \semstamp, \ksemstamp, \pmark, and \samark.

These attacks motivate reducing detector dependence on attacker-controlled sentence order and boundaries.
We therefore develop \kourstamp, which combines best-of-$N$ selection, order-robust detection, and deterministic sub-sentence units.
At the same operating point, \kourstamp lowers the attack-success rate of \eda configured for each scheme from \resultKSemEdasAsr\ on \ksemstamp to \resultKSwordEdasAsr{} at a \resultKSwordFidelityCost{}-percentage-point fidelity cost.
Within our tested models, workload, attacks, and budgets, reducing dependence on attacker-controlled text structure improves resistance to content-preserving watermark removal.

%% file: sections/appendix_extra_figures_tables.tex
\subsection{Controlled-Edit Results}
\label{app:extra-figures-tables}
\Cref{tab:isolation} shows the controlled edits are near-isolated to what they measure.
\input{figures/channel_response_appendix}

\input{tables/isolation}

%% file: figures/channel_response_appendix.tex
\begin{figure*}[!t]
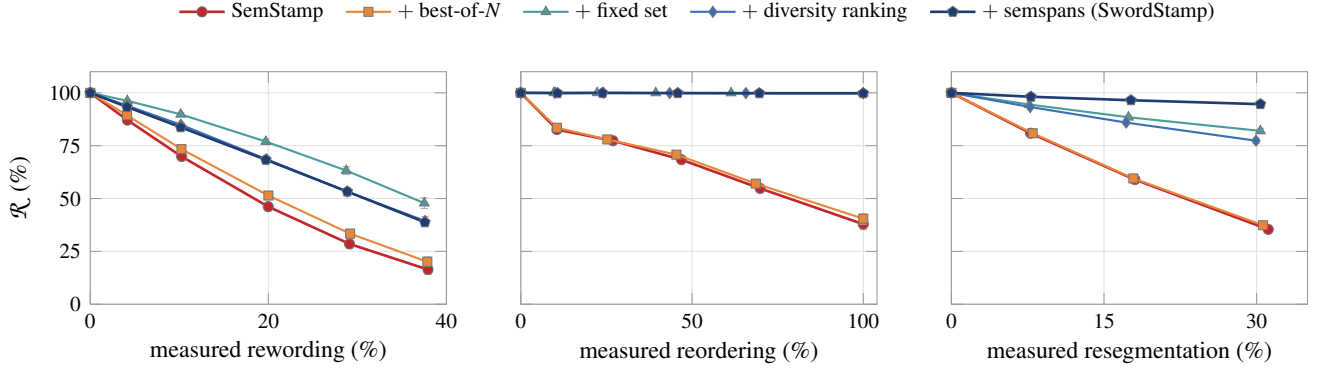

  \centering
  \chanrespbody{\altrungprefix}{\altladderbase}{\altladderfinal}
  \caption{%
    Controlled-edit response for \altladderbase\ across the five component stages of \Cref{fig:channel-response}.
    The comparison tests dependence on partition geometry.%
  }
  \label{fig:channel-response-appendix}
\end{figure*}

%% file: tables/isolation.tex
\input{tables/data/isolation}

\begin{table}[t]
  \centering
  \small
  \caption{%
    Controlled-edit measures for the \ladderbase\ base stage.
    Rows give all measures at intensity $\eta$; the target is bold.
    Across every stage, off-target values do not exceed $\isolationworstofftarget$, confirming isolation.
    Split and merge share the resegmentation measure pooled in \Cref{fig:channel-response}.%
  }
  \label{tab:isolation}
  \begin{tabular}{@{}llccc@{}}
    \toprule
    & & \multicolumn{3}{c@{}}{measured edit strength} \\
    \cmidrule(l){3-5}
    Atom & $\eta$ & reword & reorder & reseg \\
    \midrule
    \isolationrows
    \bottomrule
  \end{tabular}
\end{table}

%% file: tables/data/isolation.tex
\newcommand{\isolationrows}{%
\texttt{controlled\_reorder} & 0.10 & 0.000 & \textbf{0.109} & 0.000 \\
 & 0.25 & 0.000 & \textbf{0.269} & 0.000 \\
 & 0.50 & 0.000 & \textbf{0.492} & 0.001 \\
 & 0.75 & 0.000 & \textbf{0.729} & 0.001 \\
 & 1.00 & 0.000 & \textbf{1.000} & 0.002 \\
\addlinespace[2pt]
\texttt{synonym\_substitution} & 0.10 & \textbf{0.041} & 0.000 & 0.002 \\
 & 0.25 & \textbf{0.102} & 0.002 & 0.005 \\
 & 0.50 & \textbf{0.197} & 0.007 & 0.011 \\
 & 0.75 & \textbf{0.287} & 0.018 & 0.019 \\
 & 1.00 & \textbf{0.374} & 0.039 & 0.032 \\
\addlinespace[2pt]
\texttt{split\_midpoint} & 0.20 & 0.000 & 0.000 & \textbf{0.072} \\
 & 0.50 & 0.000 & 0.000 & \textbf{0.163} \\
 & 1.00 & 0.000 & 0.000 & \textbf{0.290} \\
\addlinespace[2pt]
\texttt{merge\_adjacent} & 0.20 & 0.000 & 0.000 & \textbf{0.089} \\
 & 0.50 & 0.000 & 0.000 & \textbf{0.197} \\
 & 1.00 & 0.000 & 0.000 & \textbf{0.339} \\
}
\newcommand{\isolationworstofftarget}{0.039}

%% file: sections/appendix_evaluation.tex
\subsection{Quality Judges and Criteria}
\label{app:eval-judge}
\label{app:eval-fidelity-criteria}

We use separate judges because clean-text fidelity and content preservation under attack are different constructs.
The \emph{reference-free fidelity judge} receives the source prompt and the generated continuation, but no reference answer.
The \emph{content-preservation judge} receives marked text $x$ as the reference and attacked text $y$ as the candidate.
The attacks neither observe nor optimize scores computed on completed outputs.

\input{tables/judge_criteria}

\textbf{System-message content.}
Both system messages assign \qwen the role of a strict, impartial judge, enumerate the four criteria defined in \Cref{tab:judge-criteria}, specify a criterion-specific scale, state the scoring rules below, and require the tagged JSON response shown afterward.
This declarative specification includes every instruction in the system messages without restating those instructions as directions to the reader.
After internal deliberation, each message requires a single line of strict JSON with an integer rating for every key and no text after the closing tag.
For fidelity, the scale is $0$ for not satisfied or empty, malformed, or unreadable text, $1$ for poor, $2$ for weak, $3$ for acceptable with noticeable issues, $4$ for strong with only minor issues, and $5$ for fully satisfied with no issues.
The fidelity message specifies that each rating concerns only the continuation and only its named criterion.
It does not reward length, detail, confidence, padding, or filler by themselves, and it does not let a weak or short source prompt excuse a weak continuation.
For content preservation, each rating measures how much of the named property is present, not whether that property is desirable: $0$ for absent, $1$ for an isolated trace, $2$ for a few instances, $3$ for a moderate degree, $4$ for most of the candidate with minor exceptions, and $5$ for fully present or present throughout the candidate.
The content-preservation message separates coverage from correctness: a wrong figure, date, or name reduces detail precision without itself counting as a contradiction, while a reversed claim or sentiment counts as a contradiction even if the candidate retains every detail.
It bases each rating only on the reference and candidate, disregards length, tone, and fluency, and assigns four zeros to an empty or unreadable candidate.

\textbf{User messages and response schema.}
For fidelity, the source prompt is the sentence recovered by the generation prompt rule from the source article's first 32 words, and the continuation excludes that prompt prefix.
The following templates preserve the literal delimiters, field order, placeholders, and required output keys.
\begin{lstlisting}[style=promptblock]
FIDELITY USER MESSAGE
[[PROMPT]]:
{source prompt}

[[CONTINUATION]]:
{watermarked continuation}

CONTENT-PRESERVATION USER MESSAGE
[[REFERENCE]]:
{marked text}

[[CANDIDATE]]:
{attacked text}

REQUIRED FINAL ANSWER FOR FIDELITY
<json>{"fluency": <0-5>, "coherence": <0-5>,
"relevance": <0-5>, "informativeness": <0-5>}</json>

REQUIRED FINAL ANSWER FOR CONTENT PRESERVATION
<json>{"content_recall": <0-5>, "detail_precision": <0-5>,
"information_injection": <0-5>, "contradiction": <0-5>}</json>
\end{lstlisting}

\textbf{Scoring and inference.}
Let each $r$ denote one criterion rating from \Cref{tab:judge-criteria}.
We give the four criteria equal weight and invert the two content-error ratings:
\begin{equation}
  \begin{aligned}
  \fid(x)
    &= 5\%\bigl(r_{\mathrm{fluency}}+r_{\mathrm{coherence}}
      +r_{\mathrm{relevance}}+r_{\mathrm{informativeness}}\bigr), \\
  \qual(x,y)
    &= 5\%\bigl(r_{\mathrm{recall}}+r_{\mathrm{detail}} \\
    &\qquad +(5-r_{\mathrm{injection}})+(5-r_{\mathrm{contradiction}})\bigr).
  \end{aligned}
  \label{eq:break-quality-aggregate}
\end{equation}
Both aggregates lie in $[0\%,100\%]$.
A score of $100\%$ requires all four fidelity properties for $\fid$, and it requires complete recall and detail precision with no information injection or contradiction for $\qual$.
We use \texttt{Qwen/Qwen3-32B} through vLLM~\cite{kwon2023vllm} with temperature $0.6$, top-$p$ $0.95$, top-$k$ $20$, minimum-$p$ $0$, and three random seeds.
We apply the model's chat template, retain only the final answer after its reasoning block, compute \Cref{eq:break-quality-aggregate} for each run, and average the three run scores for each document.
The evaluator extracts the tagged JSON object and retries an unparseable response up to three times.
It scores an empty continuation or candidate as $0$ on every criterion and excludes unparseable responses.

\textbf{Construct limits.}
Fidelity measures prompt-conditioned writing quality without comparing the continuation to a factual reference, so it does not establish factual accuracy against the source article.
Content preservation measures agreement with the marked reference, so it can preserve errors already present in that reference.
Its equal weights encode a measurement choice rather than a universal equivalence among omission, detail error, added information, and contradiction.
Both model-based scores proxy human judgments.
Neither establishes preservation of every fact and implication.

\subsection{Fidelity by Criterion}

The aggregate fidelity score can hide differences among its four criteria.
\Cref{tab:quality-criteria} reports the aggregate and each criterion for the unwatermarked baseline and all six schemes.

\input{tables/quality_criteria}

Criterion means for \ourstamp are \resultSwordCriterionGapMin--\resultSwordCriterionGapMax\ percentage points lower than for \semstamp.
Criterion means for \kourstamp are \resultKSwordCriterionGapMin--\resultKSwordCriterionGapMax\ percentage points lower than for \ksemstamp, with the largest gap in informativeness.

\subsection{\eda Prompt and Decoding}
\label{app:eval-attack-prompt}

The reported no-box \eda\ configurations use \texttt{Qwen/Qwen2.5-3B-Instruct} as the paraphraser and \texttt{BAAI/bge-base-en-v1.5} as their sole surrogate encoder.
They use the same paraphrasing prompt as Diaa et al.~\cite{diaa2024textwatermark}.
The paraphraser receives the full marked document and the following system-prompt content, reflowed only for typesetting.
\noindent{\small\color{promptAccent}\sffamily\bfseries System prompt}\par\nopagebreak
\begin{lstlisting}[style=promptblock]
You are an expert copy-editor. Please rewrite the following text in your own voice and paraphrase all sentences. Ensure that the final output contains the same information as the original text and has roughly the same length. Do not leave out any important details when rewriting in your own voice. Do not include any information that is not present in the original text. Do not respond with a greeting or any other extraneous information. Skip the preamble. Just rewrite the text directly.
\end{lstlisting}

\noindent{\small\color{promptAccent}\sffamily\bfseries User message and assistant prefix}\par\nopagebreak
\begin{lstlisting}[style=promptblock]
[[START OF TEXT]]
{marked text}
[[END OF TEXT]]

[[START OF PARAPHRASE]]
\end{lstlisting}

We apply the model's chat template to the system and user messages, then append the assistant prefix shown above.
At each output unit, the attack samples $\budget$ continuations with temperature $1.0$, top-$p$ $0.95$, and at most $96$ new tokens per candidate.
It applies the anchor rule and segmenter for the evaluated \eda\ configuration and retains the candidate with the greatest cosine distance under the BGE surrogate (\Cref{alg:edas}).
The attack conditions each subsequent sample on the full prompt and the accepted paraphrase prefix.
The sentence-level implementation stops when the paraphraser emits \eos\ or the rewrite reaches \edamaxtok\ generated tokens.

\edad\ uses the same prompt and decoding settings; \Cref{alg:detector-access} specifies how it selects candidates using the provider's region test.

\subsection{Edit Signatures of Whole-Text Attacks}
\label{app:eval-attack-channels}

\Cref{fig:attack-channels} compares rewording, reordering, resegmentation, and content preservation among outputs that evade detection at \resultPrimaryFpr\ FPR\@.
\input{figures/attack_channels}

Attack settings produce different mixtures of edits among successful evasions.
Raising \dipper's order setting from $20$ to $80$ increases mean reordering from \resultDipperEvadingReorderStart\ to \resultDipperEvadingReorderEnd, while mean rewording changes from \resultDipperEvadingRewordStart\ to \resultDipperEvadingRewordEnd.
At setting $80$, \resultDipperOrderEvadingQualityPassEnd\ of the order-sweep evasions meet $\qbar$, compared with \resultDipperLexicalEvadingQualityPassEnd\ for the lexical sweep, whose mean rewording reaches \resultDipperLexicalEvadingRewordEnd.
At the six shared budgets through $\budget=32$, \edaconfigpossent\ produces more reordering per unit of rewording than \edaconfigbagsent\ and \edaconfigbagspan.
This pattern matches the anchor objectives: positional anchors reward relocating source content, while bag anchors score each candidate against every source unit regardless of position.

Quality gating removes a larger share of evasions at higher attack settings.
Across both \dipper\ sweeps and all three \eda\ sweeps, the fraction of evasions meeting $\qbar$ decreases as diversity or budget increases.
For $\budget\in\{1,2,4,8\}$, \resultEdapEvadingQualityPassMin--\resultEdapEvadingQualityPassMax\ of \edaconfigpossent\ evasions meet $\qbar$, compared with \resultDipperEvadingQualityPassMin--\resultDipperEvadingQualityPassMax\ across \dipper's order-diversity sweep.
Because $\ASR$ requires both detector evasion and content preservation, the quality-pass difference contributes to \eda's higher $\ASR$ than \dipper on \ksemstamp\ at $\qbar=\costasrqualitybar$ in \Cref{fig:cross-scheme}.

The stage-specific \eda\ trajectories identify the structural edits that remain among evasions.
The fixed-set stage's \edaconfigbagsent\ configuration produces \resultBagSentenceResegMin--\resultBagSentenceResegMax\ mean resegmentation, compared with \resultBagSemspanResegMin--\resultBagSemspanResegMax\ for the final stage's \edaconfigbagspan\ configuration.
The final-stage configuration has a lower quality-pass fraction at \resultBagSemspanQualityLowerCount\ of the \resultBagSemspanQualityBudgetCount\ shared budgets from $4$ through $64$; at $\budget=64$, the rates are \resultBagSemspanQualityHighBudget\ and \resultBagSentenceQualityHighBudget, respectively.
This comparison describes residual evasion channels, not the effect of \semspans alone, because the stages differ and each trajectory conditions on a different set of evasions.
The trajectories therefore separate how an attack evades detection from whether the resulting text remains within the quality budget.

\subsection{Watermarked and Attacked Examples}
\label{app:eval-examples}

Individual passages expose what the two quality proxies and detector scores represent in text.
\Cref{fig:example-passages} contrasts a high-preservation evasion against \semstamp with a quality-limited evasion against \kourstamp.
\input{figures/example_passages}
In panel~(a) of \Cref{fig:example-passages}, \edap preserves \resultExampleSemQuality{} of the marked passage's content while lowering the \semstamp score below its $5\%$ FPR threshold.
The rewrite combines rewording, reordering, and resegmentation; the panel reports each measured edit rate beside the attacked text.
In panel~(b), the clean \kourstamp continuation receives \resultExampleKSwordFidelity{} fidelity, including \resultExampleKSwordFluency{} fluency.
The attack again lowers the detector score below threshold, but its \resultExampleKSwordQuality{} content-preservation score fails $\qbar=\costasrqualitybar$.
These examples illustrate the measured trade-off.

%% file: tables/judge_criteria.tex
\begin{table*}[t]
  \centering
  \small
  \setlength{\tabcolsep}{5pt}
  \caption{%
    Operational definitions and score effects for both model-based judges.
    Each criterion is rated independently from $0$ to $5$.%
  }
  \label{tab:judge-criteria}
  \begin{tabular}{@{}p{0.16\textwidth}p{0.68\textwidth}p{0.10\textwidth}@{}}
    \toprule
    Criterion & Operational definition & Score effect \\
    \midrule
    \multicolumn{3}{@{}l}{\emph{Reference-free fidelity} $\fid$} \\
    Fluency
      & The continuation uses grammatical, well-formed, and natural sentences without spelling, punctuation, or awkward-phrasing errors.
      & Raises $\fid$ \\
    Coherence
      & The ideas connect logically and remain internally consistent.
        Logical contradictions and paradoxes reduce this rating, but sentence fragments and lists do not when their ideas follow a logical sequence.
      & Raises $\fid$ \\
    Relevance
      & The text is a sensible, on-topic continuation of the supplied source prompt and remains focused rather than drifting from it.
      & Raises $\fid$ \\
    Informativeness
      & The continuation supplies substantive, specific facts or mechanisms.
        Restatements of the prompt, obvious truisms, and vague padding reduce this rating regardless of length.
      & Raises $\fid$ \\
    \addlinespace[2pt]
    \multicolumn{3}{@{}l}{\emph{Pairwise content preservation} $\qual$} \\
    Content recall
      & The candidate takes up each claim, point, and fact in the reference in some form, including a restatement, rewording, opposition, or wrong value, rather than omitting it.
      & Raises $\qual$ \\
    Detail precision
      & The specific facts, figures, names, and dates stated by the candidate are correct relative to the reference.
      & Raises $\qual$ \\
    Information injection
      & The candidate adds general claims, content, or context that the reference does not state.
      & Lowers $\qual$ \\
    Contradiction
      & The candidate asserts a claim or sentiment that directly opposes the reference's overall meaning.
      & Lowers $\qual$ \\
    \bottomrule
  \end{tabular}
\end{table*}

%% file: tables/quality_criteria.tex
\input{tables/data/quality_criteria}

\begin{table*}[t]
  \centering
  \scriptsize
  \setlength{\tabcolsep}{5pt}
  \caption{%
    Clean-text reference-free fidelity overall and by criterion.
    Cells give mean percentages and $95\%$ Student-$t$ intervals across documents after averaging three judge runs per document; higher is better.%
  }
  \label{tab:quality-criteria}
  \begin{tabular}{@{}lccccc@{}}
    \toprule
    Configuration & Fluency & Coherence & Relevance & Informativeness & Overall \\
    \midrule
    \qualitycriteriarows
    \bottomrule
  \end{tabular}
\end{table*}

%% file: tables/data/quality_criteria.tex
\newcommand{\resultSwordCriterionGapMin}{1.4}
\newcommand{\resultSwordCriterionGapMax}{2.1}
\newcommand{\resultKSwordCriterionGapMin}{3.2}
\newcommand{\resultKSwordCriterionGapMax}{6.5}
\newcommand{\qualitycriteriarows}{%
Unwatermarked & 92.2 [91.1,93.3] & 85.2 [83.6,86.8] & 84.3 [82.5,86.1] & 77.8 [76.0,79.6] & 84.9 [83.4,86.3] \\
\addlinespace[2pt]
\semstamp & 88.8 [87.5,90.0] & 79.7 [78.0,81.5] & 79.2 [77.3,81.0] & 73.9 [72.1,75.7] & 80.4 [78.9,81.9] \\
\ksemstamp & 89.2 [88.0,90.3] & 79.9 [78.2,81.5] & 78.9 [77.0,80.7] & 74.5 [72.8,76.2] & 80.6 [79.1,82.1] \\
\pmark\ (online) & 92.8 [92.0,93.6] & 85.3 [83.9,86.7] & 85.0 [83.4,86.5] & 81.1 [79.7,82.5] & 86.0 [84.8,87.2] \\
\samark & 92.1 [91.2,92.9] & 81.9 [80.4,83.4] & 77.8 [75.9,79.7] & 74.2 [72.6,75.8] & 81.5 [80.2,82.8] \\
\addlinespace[2pt]
\ourstamp & 87.4 [86.2,88.6] & 77.6 [75.9,79.3] & 77.1 [75.2,79.0] & 72.4 [70.7,74.1] & 78.6 [77.1,80.1] \\
\kourstamp & 86.0 [84.7,87.3] & 75.6 [73.9,77.3] & 75.4 [73.5,77.4] & 68.0 [66.2,69.7] & 76.2 [74.7,77.8] \\
}

%% file: figures/attack_channels.tex
\input{plots/data/attack_channels/constants}

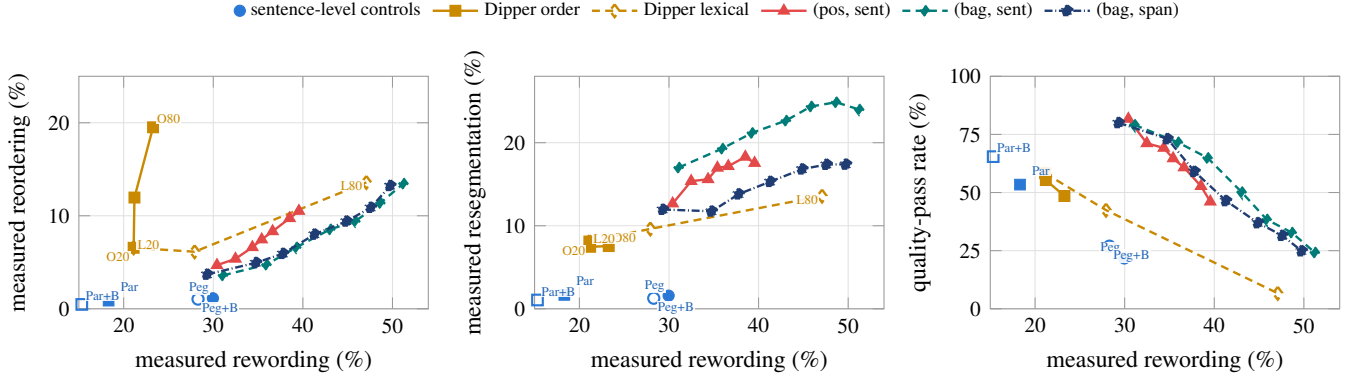
\begin{figure*}[t]
  \centering
  \begin{tikzpicture}
    \begin{groupplot}[attackchannels/signature-group]
      \nextgroupplot[attackchannels/reorder, attackchannels/signature-size, attackchannels/legend]
      \attackchannellegend
      \attackchanneltrajectories{ksemstamp}{reorder}
      \attackchanneltrajectorylabels{ksemstamp}{reorder}

      \nextgroupplot[attackchannels/reseg, attackchannels/signature-size]
      \attackchanneltrajectories{ksemstamp}{reseg}
      \attackchanneltrajectorylabels{ksemstamp}{reseg}

      \nextgroupplot[attackchannels/quality-pass, attackchannels/signature-size]
      \attackchannelqualitypasstrajectories{ksemstamp}
      \attackchannelqualitypasslabels{ksemstamp}
    \end{groupplot}
    \node[anchor=south] at ($(group c1r1.north)!0.5!(group c3r1.north)+(0,14pt)$) {\pgfplotslegendfromname{attack-channel-legend}};
  \end{tikzpicture}
  \caption{%
    Mean edit rates and quality-pass fractions among outputs that evade at \resultPrimaryFpr\ FPR\@.
    The first two panels report reordering and resegmentation against rewording; the right panel reports the fraction also meeting $\qual(x,y)\geq\qbar=\costasrqualitybar$.
    Markers identify Pegasus (circles), Parrot (squares), and bigram selection (hollow); lines connect increasing attack settings.
    Sentence-level controls and \dipper\ use \ksemstamp.
    \eda\ labels list (anchor rule, unit): \edaconfigpossent\ for \ksemstamp, \edaconfigbagsent\ for the fixed-set stage, and \edaconfigbagspan\ for the \semspan\ stage.
    Each point summarizes one attack setting over detector-evasive outputs; connected trajectories do not isolate causal effects.%
  }
  \label{fig:attack-channels}
\end{figure*}

%% file: plots/data/attack_channels/constants.tex
\newcommand{\resultDipperEvadingReorderStart}{6.6\%}
\newcommand{\resultDipperEvadingReorderEnd}{19.5\%}
\newcommand{\resultDipperEvadingRewordStart}{21.1\%}
\newcommand{\resultDipperEvadingRewordEnd}{23.3\%}
\newcommand{\resultDipperOrderEvadingQualityPassEnd}{48.5\%}
\newcommand{\resultDipperLexicalEvadingRewordEnd}{47.1\%}
\newcommand{\resultDipperLexicalEvadingQualityPassEnd}{6.7\%}
\newcommand{\resultEdapEvadingQualityPassMin}{64.7\%}
\newcommand{\resultEdapEvadingQualityPassMax}{81.7\%}
\newcommand{\resultDipperEvadingQualityPassMin}{48.5\%}
\newcommand{\resultDipperEvadingQualityPassMax}{58.3\%}
\newcommand{\resultBagSentenceResegMin}{17.0\%}
\newcommand{\resultBagSentenceResegMax}{24.9\%}
\newcommand{\resultBagSemspanResegMin}{11.7\%}
\newcommand{\resultBagSemspanResegMax}{17.4\%}
\newcommand{\resultBagSemspanQualityLowerCount}{4}
\newcommand{\resultBagSemspanQualityBudgetCount}{5}
\newcommand{\resultBagSentenceQualityHighBudget}{24.2\%}
\newcommand{\resultBagSemspanQualityHighBudget}{24.9\%}

%% file: figures/example_passages.tex
\newcommand{\resultExampleSemQuality}{96.7\%}
\newcommand{\resultExampleKSwordFidelity}{78.3\%}
\newcommand{\resultExampleKSwordFluency}{93.3\%}
\newcommand{\resultExampleKSwordQuality}{85.0\%}
\begin{figure*}[t]
\centering
\begin{minipage}{0.98\textwidth}
\footnotesize
\noindent\textbf{Units:} {\color{exampleGreen}\rule{0.8em}{0.65em}} hit; {\color{exampleYellow}\rule{0.8em}{0.65em}} hit, same region; {\color{exampleRed}\rule{0.8em}{0.65em}} miss; {\color{exampleSeed}\rule{0.8em}{0.65em}} seed. Superscripts index detector units.
\par\medskip\noindent\textbf{(a) \semstamp, \edap, $\budget=4$.}
\par\smallskip
\noindent\begin{minipage}[t]{0.485\linewidth}
\raggedright\sloppy
\textbf{Watermarked text.} $\fid(x)=90.0\%$, $z_k(x)=3.46$.\par\smallskip
{\color{exampleSeed}\textsuperscript{\scriptsize 0}Donald Trump just contradicted his own ridiculously implausible claim not to have known Matt Whitaker’s hostility toward the Russia investigation before naming him as acting attorney general.}{\color{exampleGreen}\textsuperscript{\scriptsize 1} In an interview with the Hill today, Trump said he’s “aware” that Whitaker has “spoke[n] out on it,” but went on to assert that he doesn’t necessarily believe what Whitaker says.}{\color{exampleGreen}\textsuperscript{\scriptsize 2} Or at least, Trump added, he doesn’t “believe this craziness” (emphasis mine):.}{\color{exampleGreen}\textsuperscript{\scriptsize 3} @donaldJtrumpjr: It’s “very possible” Trump Tower in Moscow was still being discussed through November 2016 @thehill https://t.co/dHrJz74pkW pic.twitter.com/p1Qq6Cf8Jd— POLITICO (@politico) December 13, 2018Trump is a consummate salesman, and today’s remarks are a textbook illustration of a technique called the “double bind.” To explain, consider a variation on Trump’s statement: “I don’t necessarily believe what Whitaker says, but I do think we should investigate him anyway.” This message actually contains two claims: One is the explicit one — namely, that Trump doesn’t necessarily believe what Whitaker says.}{\color{exampleGreen}\textsuperscript{\scriptsize 4} The second, however, is the implicit one, which says that Whitaker probably does believe what he says (since there’s no reason for Trump to think he shouldn’t be believed unless he himself believes him).}
\end{minipage}\hfill
\begin{minipage}[t]{0.485\linewidth}
\raggedright\sloppy
\textbf{Attacked text.} $\qual(x,y)=96.7\%$, $z_k(y)=1.41<\thresh(5\%)=2.12$.\par\smallskip\textit{Measured edits:} $r_{\mathrm{word}}=29.8\%$, $r_{\mathrm{order}}=13.8\%$, $r_{\mathrm{seg}}=60.0\%$.\par\smallskip
{\color{exampleSeed}\textsuperscript{\scriptsize 0}Donald Trump recently contradicted himself by denying his previous claim that he didn't know Matt Whitaker harbored hostility towards the Russia investigation.}{\color{exampleRed}\textsuperscript{\scriptsize 1} In an interview with The Hill today, Trump conceded that Whitaker had "spoken out on it," but insisted he didn’t necessarily believe what Whitaker said.}{\color{exampleGreen}\textsuperscript{\scriptsize 2} Trump even went so far as to call it “craziness.” [Politico] tweeted: @donaldJtrumpjr: It's "very possible" Trump Tower in Moscow was still being discussed through November 2016 [@thehill][https://t.co/dHrJz74pkW] pic.twitter.com/p1Qq6Cf8Jd— POLITICO (@politico) December 13, 2018.}{\color{exampleRed}\textsuperscript{\scriptsize 3} Trump is a masterful communicator, and today's statement perfectly illustrates what's known as the "double bind."}{\color{exampleRed}\textsuperscript{\scriptsize 4} To understand it, consider a modified version of his statement: "I don’t necessarily believe what Whitaker says, but I do think we should investigate him anyway."}{\color{exampleGreen}\textsuperscript{\scriptsize 5} This statement contains two distinct claims: Explicitly, Trump claims he doesn’t necessarily believe Whitaker.}{\color{exampleGreen}\textsuperscript{\scriptsize 6} Implicitly, however, he's suggesting that Whitaker probably does believe what he says (unless Trump suspects Whitaker of not being credible).}
\end{minipage}
\par\medskip\noindent\textbf{(b) \kourstamp, \edas \edaconfigbagspan, $\budget=64$.}
\par\smallskip
\noindent\begin{minipage}[t]{0.485\linewidth}
\raggedright\sloppy
\textbf{Watermarked text.} $\fid(x)=78.3\%$, $z_k(x)=6.60$.\par\smallskip
{\color{exampleRed}\textsuperscript{\scriptsize 0}A few weeks ago we showed you images of incredible}{\color{exampleRed}\textsuperscript{\scriptsize 1} robotic library logistics, including the Telelift system of the National Széchényi Library in Budapest, Hungary.}{\color{exampleGreen}\textsuperscript{\scriptsize 2} It’s easy to see why librarians are so enamored}{\color{exampleYellow}\textsuperscript{\scriptsize 3} of the system; it allows them to send automated messages to customers}{\color{exampleYellow}\textsuperscript{\scriptsize 4} for items they want and need, and delivers the materials directly to the patrons.}{\color{exampleRed}\textsuperscript{\scriptsize 5} This particular system relies on a heavily mechanized conveyor belt to bring}{\color{exampleGreen}\textsuperscript{\scriptsize 6} books to the proper patrons.}{\color{exampleGreen}\textsuperscript{\scriptsize 7} But sometimes you need more control.}{\color{exampleGreen}\textsuperscript{\scriptsize 8} Sometimes your customers are spread out over a large distance, or perhaps need access}{\color{exampleGreen}\textsuperscript{\scriptsize 9} to high-cost or fragile material.}{\color{exampleGreen}\textsuperscript{\scriptsize 10} Or maybe they’re just a little impatient.}{\color{exampleGreen}\textsuperscript{\scriptsize 11} If any of these factors is present,}{\color{exampleGreen}\textsuperscript{\scriptsize 12} then you might want to consider giving your customers}{\color{exampleRed}\textsuperscript{\scriptsize 13} mobile robots that carry their desired content to them.}{\color{exampleGreen}\textsuperscript{\scriptsize 14} In other words, you need a RoboCart.}{\color{exampleGreen}\textsuperscript{\scriptsize 15} From Bot \& Dolly.}{\color{exampleGreen}\textsuperscript{\scriptsize 16} That’s right, the company responsible for the mind-blowing film The 20th Century Foxworks production, Moonrise}{\color{exampleGreen}\textsuperscript{\scriptsize 17} Kingdom, has made a way for librarians to deliver the goods to their patrons.}{\color{exampleGreen}\textsuperscript{\scriptsize 18} Here’s the video explaining how it all works: It’s}{\color{exampleGreen}\textsuperscript{\scriptsize 19} pretty much exactly what you would expect from the guys}{\color{exampleGreen}\textsuperscript{\scriptsize 20} who made Moonrise Kingdom, but with the added bonus of actual useful content.}{\color{exampleGreen}\textsuperscript{\scriptsize 21} So now, in addition to delivering items directly to your patrons,}{\color{exampleGreen}\textsuperscript{\scriptsize 22} you can also do it with style and class, the kind}{\color{exampleGreen}\textsuperscript{\scriptsize 23} only associated with the upper class and members of the scientific community.}
\end{minipage}\hfill
\begin{minipage}[t]{0.485\linewidth}
\raggedright\sloppy
\textbf{Attacked text.} $\qual(x,y)=85.0\%$, $z_k(y)=0.88<\thresh(5\%)=2.40$.\par\smallskip\textit{Measured edits:} $r_{\mathrm{word}}=46.6\%$, $r_{\mathrm{order}}=23.7\%$, $r_{\mathrm{seg}}=21.3\%$.\par\smallskip
{\color{exampleRed}\textsuperscript{\scriptsize 0}Just a few weeks back, we introduced}{\color{exampleRed}\textsuperscript{\scriptsize 1} you to a series of impressive robotic systems in action, notably the Telelift}{\color{exampleRed}\textsuperscript{\scriptsize 2} system found in the National Széchényi Library in Budapest, Hungary.}{\color{exampleRed}\textsuperscript{\scriptsize 3} It’s hard not to be impressed by how much it’s revolutionized librarian}{\color{exampleGreen}\textsuperscript{\scriptsize 4} workflows; they can now automate notifications for customers who need}{\color{exampleGreen}\textsuperscript{\scriptsize 5} specific items, and then transport those materials right to them.}{\color{exampleYellow}\textsuperscript{\scriptsize 6} This particular Telelift relies on a robust conveyor belt to ensure books}{\color{exampleRed}\textsuperscript{\scriptsize 7} end up in the correct hands.}{\color{exampleRed}\textsuperscript{\scriptsize 8} However, there are scenarios where more personalized solutions are required.}{\color{exampleGreen}\textsuperscript{\scriptsize 9} Perhaps the library patrons are located far apart, necessitating faster}{\color{exampleGreen}\textsuperscript{\scriptsize 10} access to certain materials, or maybe the items are quite valuable}{\color{exampleRed}\textsuperscript{\scriptsize 11} and handling them manually poses a risk.}{\color{exampleRed}\textsuperscript{\scriptsize 12} In any of these cases, the Telelift might not be the best fit.}{\color{exampleRed}\textsuperscript{\scriptsize 13} For those scenarios, libraries might consider using mobile robots, or "RoboCarts"}{\color{exampleGreen}\textsuperscript{\scriptsize 14} to give patrons exactly what they want, right at their door.}{\color{exampleRed}\textsuperscript{\scriptsize 15} This innovative solution comes from the team behind The 20th Century Fox’s blockbuster}{\color{exampleRed}\textsuperscript{\scriptsize 16} Moonrise Kingdom, but don’t let that fool you—these bots also serve a purpose.}{\color{exampleRed}\textsuperscript{\scriptsize 17} This is more than just a novel method for returning and retrieving materials;}{\color{exampleGreen}\textsuperscript{\scriptsize 18} the company’s expertise also shines in the seamless integration of storytelling.}{\color{exampleRed}\textsuperscript{\scriptsize 19} So, with the Telelift system, the library doesn’t}{\color{exampleRed}\textsuperscript{\scriptsize 20} only streamline workflows, but also enhances the overall experience with a touch of cinematic flair.}
\end{minipage}
\end{minipage}
\caption{Sample watermarked and attacked passages. Colored text shows the detector's recovered units and decisions. Green and yellow both count as detector hits. Gray marks a context-dependent seed; fixed-mask detectors score the first unit.}
\label{fig:example-passages}
\end{figure*}

%% file: references.bib
@inproceedings{diaa2024textwatermark,
  title={Optimizing Adaptive Attacks against Watermarks for Language Models},
  author={Diaa, Abdulrahman and Aremu, Toluwani and Lukas, Nils},
  booktitle = {Proceedings of the 42nd International Conference on Machine Learning},
  year = {2025},
}

@inproceedings{lukas2024imagewatermark,
	title = {Leveraging Optimization for Adaptive Attacks on Image Watermarks},
	booktitle = {Proceedings of the 12th International Conference on Learning Representations},
	author = {Lukas, Nils and Diaa, Abdulrahman and Fenaux, Lucas and Kerschbaum, Florian},
  year = {2024}
}

@inproceedings{hou2024semstamp,
  title={{SemStamp}: A semantic watermark with paraphrastic robustness for text generation},
  author={Hou, Abe and Zhang, Jingyu and He, Tianxing and Wang, Yichen and Chuang, Yung-Sung and Wang, Hongwei and Shen, Lingfeng and Van Durme, Benjamin and Khashabi, Daniel and Tsvetkov, Yulia},
  booktitle={Proceedings of the 2024 Conference of the North American Chapter of the Association for Computational Linguistics: Human Language Technologies (Volume 1: Long Papers)},
  pages={4067--4082},
  year={2024}
}

@inproceedings{hou2024ksemstamp,
  title={{k-SemStamp}: A clustering-based semantic watermark for detection of machine-generated text},
  author={Hou, Abe and Zhang, Jingyu and Wang, Yichen and Khashabi, Daniel and He, Tianxing},
  booktitle={Findings of the Association for Computational Linguistics: ACL 2024},
  pages={1706--1715},
  year={2024}
}

@inproceedings{zhao2024provable,
  title={Provable Robust Watermarking for {AI}-Generated Text},
  author={Xuandong Zhao and Prabhanjan Vijendra Ananth and Lei Li and Yu-Xiang Wang},
  booktitle={The Twelfth International Conference on Learning Representations},
  year={2024},
  url={https://openreview.net/forum?id=SsmT8aO45L}
}

@inproceedings{zhao2025sok,
  title={{SoK}: Watermarking for {AI}-Generated Content},
  author={Zhao, Xuandong and Gunn, Sam and Christ, Miranda and Fairoze, Jaiden and Fabrega, Andres and Carlini, Nicholas and Garg, Sanjam and Hong, Sanghyun and Nasr, Milad and Tram{\`e}r, Florian and Jha, Somesh and Li, Lei and Wang, Yu-Xiang and Song, Dawn},
  booktitle={2025 IEEE Symposium on Security and Privacy (SP)},
  year={2025}
}

@inproceedings{kirchenbauer2023watermark,
  title={A watermark for large language models},
  author={Kirchenbauer, John and Geiping, Jonas and Wen, Yuxin and Katz, Jonathan and Miers, Ian and Goldstein, Tom},
  booktitle={International Conference on Machine Learning},
  pages={17061--17084},
  year={2023},
  organization={PMLR}
}

@misc{aaronson2023watermarking,
  title={Watermarking {GPT} outputs},
  author={Aaronson, Scott and Kirchner, Hendrik},
  year={2023},
  note={Technical report, OpenAI}
}

@article{kuditipudi2023robust,
  title={Robust distortion-free watermarks for language models},
  author={Kuditipudi, Rohith and Thickstun, John and Hashimoto, Tatsunori and Liang, Percy},
  journal={arXiv preprint arXiv:2307.15593},
  year={2023}
}

@inproceedings{charikar2002similarity,
  title={Similarity estimation techniques from rounding algorithms},
  author={Charikar, Moses S},
  booktitle={Proceedings of the Thirty-Fourth Annual ACM Symposium on Theory of Computing},
  pages={380--388},
  year={2002}
}

@inproceedings{indyk1998approximate,
  title={Approximate Nearest Neighbors: Towards Removing the Curse of Dimensionality},
  author={Indyk, Piotr and Motwani, Rajeev},
  booktitle={Proceedings of the Thirtieth Annual ACM Symposium on Theory of Computing},
  pages={604--613},
  year={1998}
}

@inproceedings{christ2024undetectable,
  title={Undetectable Watermarks for Language Models},
  author={Christ, Miranda and Gunn, Sam and Zamir, Or},
  booktitle={Conference on Learning Theory (COLT)},
  pages={1125--1139},
  year={2024},
  organization={PMLR}
}

@article{dathathri2024synthid,
  title={Scalable watermarking for identifying large language model outputs},
  author={Dathathri, Sumanth and See, Abigail and Ghaisas, Sumedh and Huang, Po-Sen and McAdam, Rob and Welbl, Johannes and Bachani, Vandana and Kaskasoli, Alex and Stanforth, Robert and Matejovicova, Tatiana and others},
  journal={Nature},
  volume={634},
  number={8035},
  pages={818--823},
  year={2024},
  publisher={Nature Publishing Group}
}

@misc{anthropic2026textwatermark,
  author={{Anthropic}},
  title={How {Claude}'s Text Watermark Works},
  year={2026},
  month=aug,
  howpublished={\url{https://www.anthropic.com/news/claude-text-watermark}},
  note={Accessed: 2026-08-20}
}

@misc{europeancommission2026transparency,
  author={{European Commission}},
  title={Code of Practice on Transparency of {AI}-Generated Content},
  year={2026},
  howpublished={\url{https://digital-strategy.ec.europa.eu/en/policies/code-practice-ai-generated-content}},
  note={Accessed: 2026-08-20}
}

@article{dabiriaghdam2025simmark,
  title={{SimMark}: A robust sentence-level similarity-based watermarking algorithm for large language models},
  author={Dabiriaghdam, Amirhossein and Wang, Lele},
  journal={arXiv preprint arXiv:2502.02787},
  year={2025}
}

@article{zhang2024personamark,
  title={{PersonaMark}: Personalized {LLM} watermarking for model protection and user attribution},
  author={Zhang, Yuehan and Lv, Peizhuo and Liu, Yinpeng and Ma, Yongqiang and Lu, Wei and Wang, Xiaofeng and Liu, Xiaozhong and Liu, Jiawei},
  journal={arXiv preprint arXiv:2409.09739},
  year={2024}
}

@article{zhang2025cohemark,
  title={{CoheMark}: A novel sentence-level watermark for enhanced text quality},
  author={Zhang, Junyan and Liu, Shuliang and Liu, Aiwei and Gao, Yubo and Li, Jungang and Gu, Xiaojie and Hu, Xuming},
  journal={arXiv preprint arXiv:2504.17309},
  year={2025}
}

@inproceedings{yu2025saemark,
  title={{SAEMark}: Steering Personalized Multilingual {LLM} Watermarks with Sparse Autoencoders},
  author={Yu, Zhuohao and Jiang, Xingru and Gu, Weizheng and Wang, Yidong and Wen, Qingsong and Zhang, Shikun and Ye, Wei},
  booktitle={Advances in Neural Information Processing Systems},
  volume={38},
  year={2025},
  doi={10.52202/085713-5301}
}

@inproceedings{huo2025pmark,
  title={{PMark}: Towards Robust and Distortion-free Semantic-level Watermarking with Channel Constraints},
  author={Huo, Jiahao and Liu, Shuliang and Wang, Bin and Zhang, Junyan and Yan, Yibo and Liu, Aiwei and Hu, Xuming and Zhou, Mingxun},
  booktitle={International Conference on Learning Representations (ICLR)},
  year={2026},
  url={https://openreview.net/forum?id=EhDgP69DJG}
}

@article{huo2026samark,
  title={{SAMark}: A Self-Anchored Text Watermarking with Paragraph-Level Paraphrase Robustness},
  author={Huo, Jiahao and Qu, Wenjie and Yan, Yibo and Zheng, Kening and Zhang, Jiaheng and Hu, Xuming and Yu, Philip S and Zhou, Mingxun},
  journal={arXiv preprint arXiv:2605.25796},
  year={2026}
}

@inproceedings{krishna2023paraphrasing,
  title={Paraphrasing evades detectors of {AI}-generated text, but retrieval is an effective defense},
  author={Krishna, Kalpesh and Song, Yixiao and Karpinska, Marzena and Wieting, John and Iyyer, Mohit},
  booktitle={Advances in Neural Information Processing Systems (NeurIPS)},
  year={2023}
}

@article{sadasivan2023aigenerated,
  title={Can {AI}-Generated Text be Reliably Detected?},
  author={Sadasivan, Vinu Sankar and Kumar, Aounon and Balasubramanian, Sriram and Wang, Wenxiao and Feizi, Soheil},
  journal={arXiv preprint arXiv:2303.11156},
  year={2023}
}

@article{chakraborty2023possibilities,
  title={On the Possibilities of {AI}-Generated Text Detection},
  author={Chakraborty, Souradip and Bedi, Amrit Singh and Zhu, Sicheng and An, Bang and Manocha, Dinesh and Huang, Furong},
  journal={arXiv preprint arXiv:2304.04736},
  year={2023}
}

@inproceedings{zhang2024watermarks,
  title={Watermarks in the Sand: Impossibility of Strong Watermarking for Generative Models},
  author={Zhang, Hanlin and Edelman, Benjamin L. and Francati, Danilo and Venturi, Daniele and Ateniese, Giuseppe and Barak, Boaz},
  booktitle={International Conference on Machine Learning (ICML)},
  year={2024}
}

@inproceedings{chen2025demark,
  title={{De-mark}: Watermark Removal in Large Language Models},
  author={Chen, Ruibo and Wu, Yihan and Guo, Junfeng and Huang, Heng},
  booktitle={Proceedings of the 42nd International Conference on Machine Learning},
  pages={9316--9333},
  volume={267},
  series={Proceedings of Machine Learning Research},
  year={2025},
  url={https://proceedings.mlr.press/v267/chen25bq.html}
}

@inproceedings{jovanovic2024watermark,
  title={Watermark Stealing in Large Language Models},
  author={Jovanovi{\'c}, Nikola and Staab, Robin and Vechev, Martin},
  booktitle={International Conference on Machine Learning (ICML)},
  year={2024}
}

@inproceedings{pang2024no,
  title={No Free Lunch in {LLM} Watermarking: Trade-offs in Watermarking Design Choices},
  author={Pang, Qi and Hu, Shengyuan and Zheng, Wenting and Smith, Virginia},
  booktitle={Advances in Neural Information Processing Systems (NeurIPS)},
  year={2024}
}

@inproceedings{mitchell2023detectgpt,
  title={{DetectGPT}: Zero-Shot Machine-Generated Text Detection using Probability Curvature},
  author={Mitchell, Eric and Lee, Yoonho and Khazatsky, Alexander and Manning, Christopher D. and Finn, Chelsea},
  booktitle={International Conference on Machine Learning (ICML)},
  year={2023}
}

@inproceedings{zhang2020pegasus,
  title={{PEGASUS}: Pre-training with Extracted Gap-sentences for Abstractive Summarization},
  author={Zhang, Jingqing and Zhao, Yao and Saleh, Mohammad and Liu, Peter J.},
  booktitle={International Conference on Machine Learning (ICML)},
  year={2020}
}

@misc{damodaran2021parrot,
  title={Parrot: Paraphrase Generation for {NLU}},
  author={Damodaran, Prithiviraj},
  year={2021},
  howpublished={\url{https://github.com/PrithivirajDamodaran/Parrot_Paraphraser}}
}

@article{grattafiori2024llama3,
  title={The {Llama} 3 Herd of Models},
  author={Grattafiori, Aaron and others},
  journal={arXiv preprint arXiv:2407.21783},
  year={2024}
}

@article{yang2025qwen3,
  title={Qwen3 Technical Report},
  author={Yang, An and others},
  journal={arXiv preprint arXiv:2505.09388},
  year={2025}
}

@article{yang2024qwen25,
  title={Qwen2.5 Technical Report},
  author={{Qwen Team}},
  journal={arXiv preprint arXiv:2412.15115},
  year={2024}
}

@article{xiao2023cpack,
  title={{C-Pack}: Packed Resources for General Chinese Embeddings},
  author={Xiao, Shitao and Liu, Zheng and Zhang, Peitian and Muennighoff, Niklas and Lian, Defu and Nie, Jian-Yun},
  journal={arXiv preprint arXiv:2309.07597},
  year={2023}
}

@inproceedings{kwon2023vllm,
  title={Efficient Memory Management for Large Language Model Serving with {PagedAttention}},
  author={Kwon, Woosuk and Li, Zhuohan and Zhuang, Siyuan and Sheng, Ying and Zheng, Lianmin and Yu, Cody Hao and Gonzalez, Joseph E. and Zhang, Hao and Stoica, Ion},
  booktitle={Proceedings of the 29th Symposium on Operating Systems Principles},
  pages={611--626},
  year={2023},
  doi={10.1145/3600006.3613165}
}

@inproceedings{reimers2019sentencebert,
  title={{Sentence-BERT}: Sentence Embeddings Using Siamese {BERT}-Networks},
  author={Reimers, Nils and Gurevych, Iryna},
  booktitle={Proceedings of the 2019 Conference on Empirical Methods in Natural Language Processing},
  pages={3982--3992},
  year={2019},
  doi={10.18653/v1/D19-1410}
}

@inproceedings{loper2002nltk,
  title={{NLTK}: The Natural Language Toolkit},
  author={Loper, Edward and Bird, Steven},
  booktitle={Proceedings of the ACL-02 Workshop on Effective Tools and Methodologies for Teaching Natural Language Processing and Computational Linguistics},
  pages={63--70},
  year={2002},
  doi={10.3115/1118108.1118117}
}

@inproceedings{zhang2020bertscore,
  title={{BERTScore}: Evaluating Text Generation with {BERT}},
  author={Zhang, Tianyi and Kishore, Varsha and Wu, Felix and Weinberger, Kilian Q. and Artzi, Yoav},
  booktitle={International Conference on Learning Representations},
  year={2020}
}

@article{raffel2020c4,
  title={Exploring the Limits of Transfer Learning with a Unified Text-to-Text Transformer},
  author={Raffel, Colin and Shazeer, Noam and Roberts, Adam and Lee, Katherine and Narang, Sharan and Matena, Michael and Zhou, Yanqi and Li, Wei and Liu, Peter J.},
  journal={Journal of Machine Learning Research},
  volume={21},
  number={140},
  pages={1--67},
  year={2020}
}

@article{heckel1978diff,
  title={A Technique for Isolating Differences Between Files},
  author={Heckel, Paul},
  journal={Communications of the ACM},
  volume={21},
  number={4},
  pages={264--268},
  year={1978},
  doi={10.1145/359460.359467}
}

@inproceedings{lin2004rouge,
  title={{ROUGE}: A Package for Automatic Evaluation of Summaries},
  author={Lin, Chin-Yew},
  booktitle={ACL Workshop on Text Summarization Branches Out},
  year={2004},
  url={https://aclanthology.org/W04-1013/}
}

@article{kendall1938rank,
  title={A New Measure of Rank Correlation},
  author={Kendall, Maurice G.},
  journal={Biometrika},
  volume={30},
  number={1--2},
  pages={81--93},
  year={1938},
  doi={10.1093/biomet/30.1-2.81}
}

@article{beeferman1999segmentation,
  title={Statistical Models for Text Segmentation},
  author={Beeferman, Doug and Berger, Adam and Lafferty, John},
  journal={Machine Learning},
  volume={34},
  number={1},
  pages={177--210},
  year={1999},
  doi={10.1023/A:1007506220214}
}

@article{pevzner2002segmentation,
  title={A Critique and Improvement of an Evaluation Metric for Text Segmentation},
  author={Pevzner, Lev and Hearst, Marti A.},
  journal={Computational Linguistics},
  volume={28},
  number={1},
  pages={19--36},
  year={2002},
  doi={10.1162/089120102317341756}
}

@article{giboulot2024watermax,
  title={{WaterMax}: Breaking the {LLM} watermark detectability-robustness-quality trade-off},
  author={Giboulot, Eva and Furon, Teddy},
  journal={Advances in Neural Information Processing Systems},
  volume={37},
  pages={18848--18881},
  year={2024}
}

@article{bahri2024blackbox,
  title={A Watermark for Black-Box Language Models},
  author={Bahri, Dara and Wieting, John and Alon, Dana and Metzler, Donald},
  journal={arXiv preprint arXiv:2410.02099},
  year={2024}
}

@article{hu2023unbiased,
  title={Unbiased Watermark for Large Language Models},
  author={Hu, Zhengmian and Chen, Lichang and Wu, Xidong and Wu, Yihan and Zhang, Hongyang and Huang, Heng},
  journal={arXiv preprint arXiv:2310.10669},
  year={2023}
}

@article{ren2023semamark,
  title={A Robust Semantics-based Watermark for Large Language Model against Paraphrasing},
  author={Ren, Jie and Xu, Han and Liu, Yiding and Cui, Yingqian and Wang, Shuaiqiang and Yin, Dawei and Tang, Jiliang},
  journal={arXiv preprint arXiv:2311.08721},
  year={2023}
}
